\documentclass[fleqn,usenatbib]{mnras}

\usepackage{newtxtext,newtxmath}
\usepackage[T1]{fontenc}

\DeclareRobustCommand{\VAN}[3]{#2}
\let\VANthebibliography\thebibliography
\def\thebibliography{\DeclareRobustCommand{\VAN}[3]{##3}\VANthebibliography}

\usepackage{graphicx}
\usepackage{amsmath}
\usepackage{enumitem}
\usepackage{booktabs}
\usepackage{tabularx}
\usepackage{xcolor}

\newcommand{\sfr}[1]{\mathrm{SFR}_{\mathrm{#1}}}
\newcommand{\Msun}{\mathrm{M}_{\odot}}
\newcommand{\Nsamp}{798}

\title[SFR indicators in H-ATLAS galaxies]{Disentangling star formation rate indicators in H-ATLAS galaxies: UV, FIR and radio diagnostics from energy-balance SED fitting}

\author[Boitumelo Gaolape et al.]{
Boitumelo Gaolape,$^{1}$
Rhodri Evans$^{1}$\thanks{E-mail: evansr@biust.ac.bw}
\\
$^{1}$Department of Physics and Astronomy, Botswana International University of Science and Technology, Palapye, Botswana\\
}

\date{Accepted XXX. Received YYY; in original form ZZZ}
\pubyear{\the\year{}}

\begin{document}
\label{firstpage}
\pagerange{\pageref{firstpage}--\pageref{lastpage}}
\maketitle

\begin{abstract}
We compare three star formation rate (SFR) tracers --- far-ultraviolet (FUV), far-infrared (FIR) and 1.4\,GHz radio emission --- for \Nsamp\ galaxies (median redshift 0.12) from the Herschel Astrophysical Terahertz Large Area Survey Data Release 1 catalogue, excluding quasars, AGN-dominated galaxies and unreliable redshifts. Energy-balance spectral energy distribution fitting with \textsc{cigale} gives their attenuations, dust luminosities and SFRs self-consistently. The FUV attenuation correlates strongly with FIR luminosity (Spearman $\rho=0.84$), largely through their common dependence on redshift (partial $\rho=0.57$ at fixed $z$), and more tightly with the infrared excess ($\rho=0.96$). The uncorrected UV SFR is over ten times below the FIR SFR for 67 per cent of the galaxies, and for all with $A_{\rm FUV}>4$\,mag. For 714 galaxies the FIRST 1.4\,GHz flux density is at the noise level and is treated as an upper limit; as detections, these values push the fits to unrealistically low dust temperatures. The 84 radio detections have a median $q_{\rm IR}$ of 2.51, close to the assumed 2.58. SFRs averaged over 10 and 100\,Myr differ by under 2 per cent, because the model star formation histories are smooth. Without access to high-performance computing, all galaxies were fitted at $z=0$, with distance-dependent quantities rescaled. This lowers dust temperatures by a factor of about $(1+z)$ and biases stellar masses and, at $z>0.2$, attenuations, but changes dust luminosities and SFRs by only a few hundredths of a dex. These fits, with 16.3 million models per galaxy, show what UV-to-radio SED modelling can deliver on modest hardware.
\end{abstract}

\begin{keywords}
galaxies: star formation -- galaxies: photometry -- infrared: galaxies -- radio continuum: galaxies -- ultraviolet: galaxies
\end{keywords}

\section{Introduction}
\label{sec:intro}

The star formation rate (SFR) describes how galaxies convert gas into stars, and measuring it is central to understanding galaxy evolution. In cosmological models the SFR of a galaxy is shaped by its history of gas cooling, star formation, chemical enrichment, and feedback from supernovae and active nuclei, as well as by mergers, interactions and gas accretion \citep{lozano2023model}. Because low-mass stars leave only weak signatures in the integrated light of a galaxy, SFRs are inferred indirectly, either from emission that is dominated by massive, short-lived stars \citep{dutkowska2023star} or from stellar population synthesis modelling of the full SED \citep{yu2024exploring}.

Several observational tracers are in common use, each sensitive to a different physical process. Ultraviolet (UV) emission comes directly from the photospheres of young, massive stars. Far-infrared (FIR) emission comes from dust heated by those stars, so that the total infrared luminosity, $L_{\rm IR}$ (8--1000\,$\mu$m), measures the stellar light that has been absorbed and re-emitted by dust \citep{battisti2017ultraviolet}. Radio continuum emission at 1.4\,GHz is dominated by non-thermal synchrotron radiation from cosmic-ray electrons accelerated in the supernova remnants of massive stars \citep{condon1992radio, werhahn2021cosmic}, with a smaller contribution from thermal free--free emission in H\,\textsc{ii} regions. The three tracers are not equivalent. The UV is strongly attenuated by dust. The FIR depends on how much of the stellar light is absorbed, on the dust geometry and on non-stellar heating sources, and the radio can be contaminated by active galactic nuclei (AGN). They also respond to star formation over different timescales: very roughly, from $\sim 10$\,Myr for the emission dominated by the youngest stars to $\sim 100$\,Myr for the UV, FIR and radio emission in galaxies with extended star formation histories \citep{kennicutt1998star, KennicuttEvans2012}, with the exact sensitivity depending on the star formation history (SFH).

The infrared and UV are complementary. Since star formation takes place in dense molecular clouds, much of the radiation from young stars is absorbed by the surrounding dust, preferentially at short wavelengths, and re-emitted mainly between $\sim 10$ and 120\,$\mu$m \citep{Ot_Floranes_2010}. When the colour excess $E(B-V)$\footnote{$E(B-V)$ is the colour excess, which quantifies the amount of reddening due to interstellar dust. Higher values correspond to greater dust attenuation.} exceeds $\sim 0.5$, almost all of the UV and optical radiation of the young stars is absorbed and re-emitted in the FIR \citep{Ot_Floranes_2010}, and the FIR luminosity is then a good measure of the total energy output of massive stars \citep{kennicutt1998star}. The sensitivity and angular resolution of the \textit{Spitzer Space Telescope} \citep{werner2004spitzer} increased the use of infrared SFR indicators \citep{shi2013star}. For the UV, the usual approach is to correct for attenuation by using the relation between the infrared excess, $\mathrm{IRX}=L_{\rm IR}/L_{\rm UV}$, and the UV spectral slope $\beta$ (the IRX--$\beta$ relation; \citealt{Meurer1999}), which converts the observed slope into an attenuation \citep{boquien2016towards}. This works well for starbursts but less well for more quiescent galaxies, in part because the attenuation law varies between galaxies \citep{SalimNarayanan2020}. A different strategy is to use hybrid UV+IR estimators, which correct the UV using dust emission in a specific infrared band; \citet{boquien2016towards} showed that the performance of these estimators depends on the specific SFR at 24\,$\mu$m and in the total infrared, and on the stellar mass surface density at 70 and 100\,$\mu$m.

The FIR and radio emission of star-forming galaxies are linked by the FIR--radio correlation (FIRC), which was characterised by \citet{condon1992radio} and holds over several decades in luminosity. It is thought to arise because the same massive stars both heat the dust and, through their supernovae, accelerate the cosmic rays that produce the synchrotron emission \citep{magnelli2015far}.

Given these differences, it is important to compare the tracers under a single, uniform modelling framework, so that differences between them can be attributed to the physics and not to differences in method. In this paper we use energy-balance SED fitting with \textsc{cigale} (Code Investigating GALaxy Emission; \citealt{boquien2019cigale}) to study the relative behaviour of UV, FIR and radio SFR diagnostics for galaxies in H-ATLAS DR1. Our aims are to (i) examine how the FUV and NUV attenuation relate to the FIR emission; (ii) test how the FIR and radio emission are related, and to what extent that relation is independent of the modelling; (iii) compare SFRs averaged over different timescales; and (iv) identify when and why the indicators disagree, and what this implies for multiwavelength SFR measurements in large surveys. The comparison of \textsc{cigale} modules on which our choice of models is based is presented in a companion paper \citep{evans_etal2026}.

The fits are based on the MSc Thesis of the first author, which used \textsc{cigale} v2022.1 on a laptop with 8 cores and 8\,GB of memory belonging to the Botswana International University of Science and Technology (BIUST) and lent to the first author by the second; there is no high-performance computing (HPC) facility in Botswana, a situation shared with many other countries in Africa and in other parts of the developing world. The final fits in this paper use the same version of the code and the same photometry, with three changes made after the Thesis: we removed from the sample the sources with unreliable redshifts, the quasars and the galaxies whose mid-infrared emission is dominated by an AGN (Section~\ref{sec:data}); the parameter grid of the Thesis (9\,072\,000 models per galaxy) was extended by us (in particular to lower and higher dust temperatures); and the faint 1.4\,GHz flux densities were entered as upper limits (Sections~\ref{sec:data} and~\ref{sec:cigale}). The paper has two purposes. The first is to report the main results of that work. The second is to show what can be achieved with energy-balance SED fitting of several hundred galaxies from the UV to the radio on a modest computer, which is the situation of many astronomers outside the institutions with HPC access, and where the limits of such modelling lie. Sections~\ref{sec:cigale} and~\ref{sec:discussion} describe the computational constraints and their consequences.

The paper is organised as follows. Section~\ref{sec:data} describes the data and sample. Section~\ref{sec:cigale} describes the SED modelling, including our treatment of redshift. Section~\ref{sec:indicators} defines the SFR indicators. Section~\ref{sec:results} presents the results, Section~\ref{sec:discussion} discusses them and their limitations, and Section~\ref{sec:conclusions} gives our conclusions.

\section{Data and sample}
\label{sec:data}

\subsection{The H-ATLAS DR1 catalogue}

H-ATLAS \citep{eales2010herschel} is a wide-area extragalactic survey made with the \textit{Herschel Space Observatory}. It was allocated 600\,hours of observing time and covers $\sim 600$\,deg$^2$ in five photometric bands, using the PACS camera at 100 and 160\,$\mu$m and the SPIRE camera at 250, 350 and 500\,$\mu$m. We use the first data release (DR1; \citealt{valiante2016herschel}), with optical counterparts from \citet{bourne2016herschel}. DR1 comprises five fields: a $\sim 170$\,deg$^2$ field near the North Galactic Pole, observed by the Sloan Digital Sky Survey (SDSS); a $\sim 285$\,deg$^2$ field near the South Galactic Pole, observed by the 2-Degree-Field (2dF) Galaxy Redshift Survey; and three fields of $\sim 54$\,deg$^2$ each on the celestial equator, observed by SDSS, 2dF and the Galaxy And Mass Assembly (GAMA) survey \citep{valiante2016herschel}. The fields were chosen to have low Galactic cirrus emission, as shown by the IRAS 100\,$\mu$m maps, and good multiwavelength coverage. All the galaxies in our sample lie in the equatorial GAMA 9-h field (G09; Section~\ref{sec:selection}).

Because of the strongly negative $k$-correction at sub-millimetre wavelengths, H-ATLAS detects dusty galaxies at both low redshift ($z \lesssim 0.1$) and high redshift ($z>1$) \citep{valiante2016herschel}. The sample is selected from SPIRE detections and is therefore biased towards FIR-bright, dusty galaxies; we return to this in Section~\ref{sec:discussion}.

\subsection{Multiwavelength photometry and sample selection}
\label{sec:selection}

Our sample consists of \Nsamp\ galaxies from DR1, selected as described below. The photometry comes from the H-ATLAS DR1 catalogue \citep{valiante2016herschel,bourne2016herschel}, which provides the \textit{Herschel} SPIRE and PACS flux densities and the matched-aperture GALEX \citep{Martin2005}, SDSS \citep{York2000}, VISTA (VIKING; \citealt{Edge2013}) and WISE \citep{Wright2010} photometry from the GAMA survey (the `GAMA panchromatic photometry' columns of the catalogue; \citealt{Driver2016}), and to which we added 1.4\,GHz flux densities from the FIRST survey \citep{Becker1995}. The bands are:
\begin{itemize}
    \item GALEX FUV (1528\,\AA) and NUV (2271\,\AA);
    \item SDSS $ugriz$;
    \item VISTA $ZYJHK_s$;
    \item WISE W1--W4 (3.4, 4.6, 12 and 22\,$\mu$m);
    \item \textit{Herschel} PACS 100 and 160\,$\mu$m, and SPIRE 250, 350 and 500\,$\mu$m;
    \item 1.4\,GHz radio flux density from FIRST.
\end{itemize}
All the flux densities in the DR1 catalogue are in Jy; we converted them to mJy, the unit used by \textsc{cigale}.
The \textsc{cigale} run log (Fig.~\ref{fig:general info on models}) shows that 22 photometric bands were fitted: GALEX (2), SDSS (5), VISTA (5), WISE (4), \textit{Herschel} (5) and the VLA L-band 1.4\,GHz flux density (1). The radio flux density is therefore an input to the fit, and the \texttt{radio} module relates it to the infrared (Section~\ref{sec:radio_module}).

The coverage is not complete (Table~\ref{tab:coverage}). Every galaxy has SPIRE photometry, optical and near-infrared photometry, and a radio flux density (a measurement for 84 galaxies and an upper limit for 714; see below), but 20 per cent (160 galaxies) have no GALEX photometry in either band, and 195 (24 per cent) have no FUV measurement. W4 is missing for 148 galaxies, and PACS for 35 (the sources outside the PACS maps; Section~5.3 of the DR1 README). A missing measurement is given the flag value expected by \textsc{cigale}, and the band is not used in the fit of that galaxy. 476 galaxies (60 per cent) have all 22 bands, 619 (78 per cent) have at least one band in each of the seven groups of Table~\ref{tab:coverage}, and the median number of bands fitted per galaxy is 22. The consequence for this paper is that for the 160 galaxies without GALEX data, the FUV attenuation and the model FUV flux are inferred from the other bands and are not constrained by an FUV measurement. The median fitted $A_{\rm FUV}$ is 4.1\,mag for these 160 galaxies, compared with 2.4\,mag for the 603 galaxies with an FUV measurement, so the results involving $A_{\rm FUV}$ in Section~\ref{sec:results} should be read with this in mind. The galaxies without an FUV measurement are at higher redshifts (median $z=0.19$ for the 160 without GALEX data and 0.26 for the 35 with an NUV measurement only, compared with 0.11) and have higher dust luminosities, which suggests that most of them are fainter and more attenuated galaxies and not galaxies outside the GALEX coverage, although we have not checked the coverage directly. Restricting the comparisons of Section~\ref{sec:comparison} to the 603 galaxies with an FUV measurement does not change the results: in the four bins of Table~\ref{tab:sfr_ratios} (209, 241, 104 and 49 galaxies) the median $\sfr{UV,corr}/\sfr{100}$ is 1.21, 1.13, 1.16 and 1.24, the median $\sfr{FIR}/\sfr{100}$ is 1.15, 1.61, 1.83 and 2.07, and the fraction with $\sfr{UV}/\sfr{FIR}<0.1$ is 0.08, 0.88, 1.00 and 1.00, and the Spearman coefficient between $A_{\rm FUV}$ and $\sfr{UV}/\sfr{FIR}$ is $-0.95$ ($-0.96$ for all \Nsamp\ galaxies).

\begin{table}
	\caption{Number of the \Nsamp\ galaxies with a measured flux density in each band of the \textsc{cigale} fits. The last two columns give the number of galaxies with at least one, and with all, of the bands in each group. $^{a}$84 measurements and 714 upper limits.}
	\label{tab:coverage}
	\centering
	\small
	\begin{tabular}{@{}llrrr@{}}
		\toprule
		Group & Band & $N$ & $\geq1$ band & All bands \\
		\midrule
		GALEX & FUV & 603 & 638 & 603 \\
		      & NUV & 638 &     &     \\
		SDSS  & $u,g,r,i,z$ & 798 each & 798 & 798 \\
		VISTA & $Z,Y,J,K_s$ & 798 each & 798 & 785 \\
		      & $H$ & 785 &     &     \\
		WISE  & W1, W2 & 792 each & 792 & 649 \\
		      & W3 & 787 &     &     \\
		      & W4 & 650 &     &     \\
		\textit{Herschel} PACS & 100\,$\mu$m & 763 & 763 & 762 \\
		      & 160\,$\mu$m & 762 &     &     \\
		\textit{Herschel} SPIRE & 250, 350, 500\,$\mu$m & 798 each & 798 & 798 \\
		Radio & FIRST 1.4\,GHz$^{a}$ & 798 & 798 & 798 \\
		\bottomrule
	\end{tabular}
\end{table}

\emph{Sample selection.} The sample was defined by the manual look-up of the radio flux densities. B.~Gaolape worked through the DR1 catalogue (120\,230 sources) in the order in which the sources appear in it, and looked up their FIRST 1.4\,GHz flux densities on the FIRST website. Within the first 6\,373 rows of the catalogue she recorded a FIRST entry for 1\,031 sources, and stopped there because the look-up was done by hand. Of these, 852 were fitted in the Thesis. The other 179 were not fitted: 162 have no spectroscopic redshift in DR1 (\texttt{Z\_QUAL}\,=\,$-99$; these include the seven sources with \texttt{GSQ\_FLAG}\,=\,3), and 15 of the remaining 17 have no GAMA counterpart and hence no UV-to-mid-infrared photometry. Of the other two, G09.DR1.857 has a catalogue redshift of $-4\times10^{-5}$, for which no distance can be calculated, and G09.DR1.770 was omitted by an oversight when the look-up was transferred to the input file; it is a galaxy at $z=0.116$ with a FIRST flux density at the noise level, and its omission does not affect the results. The Thesis sample was therefore selected on the existence of a spectroscopic redshift, of any quality, and not on \texttt{Z\_QUAL}\,$\geq3$. All of the 852 lie in the G09 field, which comes first in the catalogue, and they are drawn from its first 6\,373 rows (of 39\,660). Within each field the rows of the catalogue \citep{valiante2016herschel} follow the numbering of the DR1 names (G09.DR1.1, G09.DR1.2, \ldots), which runs approximately, but not strictly, in order of decreasing 250-$\mu$m signal-to-noise ratio: the Spearman coefficient between row number and $S_{250}/\sigma_{250}$ is $-0.96$ in each field, against $-0.88$ for $S_{250}$ itself. The first rows of G09 therefore contain the most significant, and mostly the brightest, 250-$\mu$m sources in the field. The selection is therefore in effect a 250-$\mu$m flux-density limit in one field and not a random draw from DR1. Compared with the DR1 galaxies that have a reliable redshift and at least eight bands detected at $\geq3\sigma$, the 852 have a median $S_{250}$ of $\approx100$\,mJy (41\,mJy for DR1) and a median redshift of 0.13 (0.22), and 57 per cent of them are detected at 100\,$\mu$m (12 per cent for DR1).

For this paper we removed five groups of sources from the Thesis sample: the 13 with an unreliable redshift (\texttt{Z\_QUAL}\,$<3$); the 11 classified as quasars in DR1 (\texttt{GSQ\_FLAG}\,=\,2), whose emission cannot be represented by our models, which have no AGN component (Section~\ref{sec:agn}); and 22 galaxies whose WISE colours indicate that an AGN dominates the mid-infrared emission. We define these as galaxies with $W1-W2\geq0.8$ (Vega) and a signal-to-noise ratio of at least 5 in both bands \citep{stern2012mid}. We calculated the colours from the GAMA flux densities, which are on the AB system, with AB-to-Vega offsets of 2.699\,mag (W1) and 3.339\,mag (W2) \citep{jarrett2011spitzer}. This criterion was calibrated with profile-fit WISE magnitudes, whereas we use the matched-aperture GAMA photometry, which can give different colours for extended galaxies. We checked the classification against the profile-fit photometry of the AllWISE catalogue \citep{Wright2010,cutri2013allwise}, matched within 3\,arcsec of the optical positions, for the 22 excluded galaxies and all \Nsamp\ retained ones; all the excluded galaxies and 789 of the retained ones have a counterpart. Twenty of the 22 excluded galaxies also have $W1-W2\geq0.8$ in AllWISE, and the other two have 0.78. Of the retained galaxies, three have $W1-W2=0.84$--0.87 in AllWISE, all of them within 0.03\,mag of the limit in the GAMA photometry; none of the others, including the 14 whose GAMA colours do not meet the signal-to-noise requirement, satisfies the criterion in AllWISE. The classification is therefore the same for 806 of the 811 galaxies with a counterpart, and the five that differ all lie within 0.1\,mag of the limit, comparable to the scatter between the two sets of colours (robust standard deviation 0.07\,mag, with AllWISE redder by a median of 0.04\,mag). The individual magnitudes also agree: for the galaxies that are unresolved in AllWISE and detected at a signal-to-noise ratio of at least 10 (551 in W1 and 547 in W2), the GAMA W1 and W2 magnitudes are brighter than the profile-fit ones by medians of 0.15 and 0.10\,mag, as expected for aperture photometry of slightly resolved galaxies, which confirms the flux units and the zero-points. We kept the selection based on the GAMA photometry, which is the photometry used in the fits. The criterion selects galaxies in which an AGN dominates the mid-infrared; it is incomplete for low-luminosity and composite AGN, and it does not identify AGN that are bright only in the radio, so the remaining sample is not free of AGN (Sections~\ref{sec:agn} and~\ref{sec:discussion}). Finally, we removed the six remaining galaxies that have no GAMA counterpart, for which only the \textit{Herschel} and 1.4\,GHz bands are available, so that the stellar emission and the attenuation are not constrained, and the two galaxies (G09.DR1.6230 and G09.DR1.6250) for which the FIRST look-up gave more than one component, so that the 1.4\,GHz flux density of the galaxy is ambiguous. The five groups do not overlap, and the final sample has \Nsamp\ galaxies. All seven sources of the Thesis sample with $z>1$ were quasars, so the highest redshift in the final sample is $z=0.74$; its median redshift is 0.12 (quartiles 0.06 and 0.23), and its median $S_{250}$ is 101\,mJy. Of the \Nsamp\ galaxies, 84 have a FIRST flux density of at least 0.5\,mJy and 714 a lower value. The cuts do not change the main results: with the 54 excluded galaxies included (8, 15, 7 and 24 of them in the four $A_{\rm FUV}$ bins of Table~\ref{tab:sfr_ratios}), the Spearman coefficient between $A_{\rm FUV}$ and $L_{\rm dust}$ is 0.80 instead of 0.84, the uncorrected UV SFR is below a tenth of the FIR SFR for 68 instead of 67 per cent of the galaxies, and the median ratios of the corrected UV and FIR SFRs to $\sfr{100}$ are 1.20 and 1.70 instead of 1.20 and 1.72. It is representative of FIR-bright, low-redshift galaxies in G09 and not of DR1 as a whole, and it is not complete in any band other than SPIRE.

\emph{Radio flux densities.} The 1.4\,GHz flux densities were looked up on the FIRST website for each galaxy by B.~Gaolape, and an uncertainty of 15 per cent was assumed. The distribution is concentrated at the faint end: 641 of the \Nsamp\ galaxies (80 per cent) have $S_{1.4}\leq0.16$\,mJy, with a sharp peak at 0.14--0.155\,mJy, and only 73 (9 per cent) are above 1\,mJy, about the detection limit of the FIRST catalogue. The faint values are therefore at the level of the FIRST noise ($\approx0.15$\,mJy) and are not detections, and a 15 per cent uncertainty understates their true uncertainty, which is of the order of the value itself. The NVSS values tabulated for the same galaxies support this reading: for all 714 galaxies with a FIRST value below 0.5\,mJy the NVSS value is consistent with zero (the largest is 0.07 in the units of the tabulated column), so there is no sign of extended emission missed by FIRST, whereas for the 37 of the remaining 84 galaxies that have an NVSS value, the median ratio of the FIRST to the NVSS flux density is 0.79. We therefore treat the 714 galaxies with $S_{1.4}<0.5$\,mJy as non-detections. In the \textsc{cigale} input they are upper limits of 0.75\,mJy (five times the noise), with a scale of 0.15\,mJy: \textsc{cigale} then penalises models whose radio flux density exceeds the limit and does not penalise models below it. The 84 galaxies above 0.5\,mJy keep their measured flux densities and the 15 per cent uncertainty. In the first version of the fits all the values were entered as detections; Section~\ref{sec:radio} shows how much this changes the results. The radio comparison with observations (the observed $q_{\rm IR}$) is therefore possible only for the 84 galaxies with measured flux densities.

\subsection{Redshifts and morphology}

\begin{figure}
	\includegraphics[width=\columnwidth]{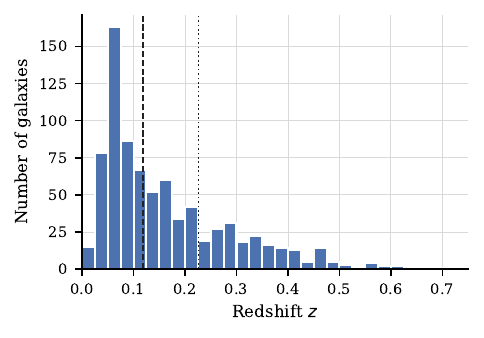}
	\caption{The redshift distribution of the \Nsamp\ galaxies, in bins of 0.025. The dashed and dotted lines mark the median ($z=0.12$) and the third quartile ($z=0.23$); the highest redshift is $z=0.74$.}
	\label{fig:redshift_histogram}
	\par\smallskip{\small Alt text: Histogram of redshift from 0 to 0.75 in bins of 0.025. Most galaxies lie below $z=0.3$, with a peak of about 160 galaxies near $z=0.05$; vertical dashed and dotted lines mark the median, 0.12, and the upper quartile, 0.23.}
\end{figure}

The redshift distribution is shown in Fig.~\ref{fig:redshift_histogram}. It spans $z=0.011$--$0.74$, with a median of $z=0.12$ and quartiles at $z=0.06$ and $0.23$, so the sample is dominated by objects in the local Universe. Morphological classifications from Galaxy Zoo \citep{lintott2011galaxy} are available for only 98 of the \Nsamp\ galaxies: 60 are classified as spirals, one as elliptical and the other 37 as uncertain. The sample as a whole therefore cannot be described as a sample of spiral galaxies, and in what follows we refer to `galaxies' unless we are referring to the Galaxy Zoo subsample.

\section{SED modelling with \textsc{cigale}}
\label{sec:cigale}

\subsection{Method}

We use \textsc{cigale} (v2022.1; \citealt{boquien2019cigale}) to fit the photometry of each galaxy. \textsc{cigale} builds a grid of models from user-specified modules, computes the stellar emission and the fraction of the UV/optical energy absorbed by dust, and re-emits that energy in the infrared (energy balance), thereby linking the UV/optical and FIR parts of the SED. The model fluxes are then compared with the observed fluxes, and physical properties are estimated from the fit. The quantities used in this paper are those of the best-fitting model of each galaxy (the `best' columns of the \textsc{cigale} output), and not the likelihood-weighted means of the \texttt{pdf\_analysis} module.

The modules and models we use are listed in Table~\ref{tab:modules}, and the parameter grid is given in Table~\ref{tab:cigale_parameters} in Appendix~\ref{app:cigale}. The grid contains $8\times 9\times 3 \times 10 \times 7 \times 3 \times 12 \times 5 \times 6 = 16\,329\,600$ models per galaxy (the numbers of values of $\tau$, age, metallicity, \texttt{Av\_young}, \texttt{Av\_old\_factor}, the UV bump amplitude, the dust temperature, $\beta$ and $\alpha$). The choice of each module followed extensive testing, which is described in \citet{evans_etal2026}; here we summarise the models and the reasons for the choices.

\emph{Computational constraints.} The fits of the Thesis were run in 2024 with \textsc{cigale} v2022.1 (the later releases v2025.0 and v2025.1 were not yet available) on a BIUST laptop lent to the first author (HP ProBook 445 G9, with an eight-core AMD Ryzen 7 5825U processor, 8\,GB of DDR4-3200 memory in a single module and a 512\,GB NVMe solid-state disk), with four cores in use (\texttt{cores\,=\,4} in \texttt{pcigale.ini}). In tests, runs with more than about $1.5$--$1.8\times10^{7}$ models per galaxy failed for lack of memory, and the grid of the Thesis, which has 9\,072\,000 models per galaxy, was designed to stay below this limit. In the Thesis grid the dust temperatures were 21--40\,K (five values), $\tau\leq8000$\,Myr and the ages up to 4\,Gyr; for 81 per cent of the galaxies the best-fitting dust temperature was the lowest value, 21\,K, and many other parameters piled up at the edges of their ranges. The grid used in this paper is not the grid of the Thesis. After the Thesis was completed we extended the ranges of the parameters and changed some of the values (Table~\ref{tab:cigale_parameters}), to reduce this pile-up. This was done in two steps. An intermediate grid of 12\,859\,560 models (dust temperatures 12.5--40\,K) removed the pile-up at the lowest temperature, but a large fraction of the best-fitting models still lay at the upper edge of the grid for the dust temperature (11 per cent), the emissivity index $\alpha$ (14 per cent) and $\tau$ (21 per cent). The final grid, used for all results in this paper unless stated otherwise, was designed to reduce these pile-ups: we extended the ranges of $T$, $\alpha$, $\tau$ and the ages, added the value \texttt{Av\_old\_factor}\,=\,1 (equal attenuation of the old and young stars, which is the physical maximum), reduced the number of values of $\beta$ from seven to five, and dropped values that were seldom chosen. It has 16\,329\,600 models per galaxy, within the memory limit of the laptop used for the Thesis. The final fits with this grid were run with the same version of the code on a Linux mini PC owned by the second author (Minisforum UM790 Pro, with an eight-core, 16-thread AMD Ryzen 9 7940HS processor, 32\,GB of DDR5-5600 memory and a 1\,TB NVMe solid-state disk), using six cores and dividing the grid into 20 blocks, and took 1\,h\,38\,min for the 852 galaxies of the Thesis sample (Fig.~\ref{fig:general info on models}); the results for the \Nsamp\ galaxies of this paper are taken from that fit, since the fit of each galaxy is independent of the others. The design of the grid is still a compromise between the number of values of each parameter and the number of parameters that can be varied: several parameters are sampled at only a few values, and for a fraction of the galaxies the best-fitting model still lies at the edge of the grid (Section~\ref{sec:discussion}; the remaining cases are mostly physical limits or degeneracies, not limits that more grid points would remove). The choice of modules and the design of the grid were guided by tests on 50 of the 852 galaxies, all with complete photometry from the FUV to 500\,$\mu$m and drawn to follow the redshift distribution of the galaxies with complete photometry, with 113 configurations of 1 to $1.7\times10^{7}$ models; the tests are described in \citet{evans_etal2026}, and their results for the dust-emission module are summarised in Section~\ref{sec:dust_module}.

\begin{table}
	\caption{Models used for each \textsc{cigale} module.}
	\label{tab:modules}
	\centering
	\begin{tabular}{ll}
		\toprule
		Module & Model used \\
		\midrule
		SFH & \texttt{sfhdelayed} \\
		SSP & \texttt{bc03} \\
		Nebular emission & \texttt{nebular} \\
		Dust attenuation & \texttt{dustatt\_powerlaw} \\
		Dust emission & \texttt{casey2012} \\
		AGN & not included (AGN excluded from the sample) \\
		X-ray & not included \\
		Radio & \texttt{radio} \\
		Rest-frame parameters & not included \\
		Redshift + IGM & \texttt{redshifting} \\
		\bottomrule
	\end{tabular}
\end{table}

\subsection{Star formation history}

\textsc{cigale} provides three analytical SFH modules: single or double exponentials (\texttt{sfh2exp}), a delayed SFH with an optional exponential burst (\texttt{sfhdelayed}), and periodic star formation (\texttt{sfhperiodic}) \citep{boquien2019cigale}. We adopt \texttt{sfhdelayed}, in which the SFR rises nearly linearly, peaks at $t=\tau$ and then declines exponentially,
\begin{equation}
	\mathrm{SFR}(t) \propto \frac{t}{\tau^{2}}\, e^{-t/\tau},
	\label{eq:sfh}
\end{equation}
which gives a smooth variation that is more realistic for real galaxies than abrupt models \citep{boquien2019cigale}. Our grid has $\tau$ between 100 and 8000\,Myr and ages between 500 and 10\,000\,Myr, and no burst is included (\texttt{f\_burst}\,=\,0). Because the fits are made at $z=0$ (Section~\ref{sec:redshift}), ages up to 10\,Gyr are allowed for all galaxies, although they exceed the age of the Universe for galaxies at $z\gtrsim0.3$.

\subsection{Stellar populations and nebular emission}

The intrinsic stellar emission is computed with single stellar population (SSP) libraries. \textsc{cigale} offers the models of \citet{maraston2005evolutionary} (\texttt{m2005}) and \citet{bruzual2003stellar} (\texttt{bc03}), and the latter gave the best fits to our galaxies. We use the \citet{salpeter1955luminosity} initial mass function (IMF), which fitted our galaxies as well as that of \citet{chabrier2003galactic} (in tests on 50 of them the reduced $\chi^2$ changed by a median of about 1 per cent; \citealt{evans_etal2026}), with metallicities of 0.008, 0.02 and 0.05. Following \citet{charlot2000simple}, the stellar populations are first treated as dust-free, with young ($<10$\,Myr) and old stars computed separately so that age-dependent attenuation can be applied by the dust attenuation module.

A large part of the light of the hottest stars is emitted in the Lyman continuum ($\lambda<912$\,\AA), where photons ionise hydrogen. The surrounding gas re-emits part of this energy as continuum (free--free, free--bound and two-photon), reaching into the radio regime, and as emission lines. Nebular emission is often neglected because its contribution to broad-band fluxes is small in quiescent galaxies, but it is a useful probe of recent star formation. \textsc{cigale} uses pre-computed templates based on \textsc{cloudy} \citep{inoue2011rest, ferland1998cloudy, ferland20132013}, and we include them so that the continuum and line contributions are present in the SEDs \citep{boquien2019cigale}.

\subsection{Dust attenuation}

Starlight is absorbed and scattered by dust along the line of sight, which is known as dust attenuation \citep{Markov2024}. It is more complicated than extinction, since it depends on the (often unknown) geometry of stars and dust \citep{Conroy_2013}. \textsc{cigale} allows the attenuation law to be the \citet{calzetti2000dust} law or a power law,
\begin{equation}
	k(\lambda)_{\rm PL} \propto \lambda^{\alpha},
	\label{eq:powerlaw}
\end{equation}
as in \citet{charlot2000simple} and \citet{salim2016galex}, where $\lambda$ is the wavelength and $\alpha$ is the exponent that sets the shape of the curve; a more negative $\alpha$ means a steeper curve, with stronger attenuation at shorter wavelengths. After testing the available models (\texttt{dustatt\_powerlaw}, \texttt{dustatt\_calzleit}, \texttt{dustatt\_modified\_CF00}, \texttt{dustatt\_modified\_starburst} and \texttt{dustatt\_2powerlaws}; \citealt{evans_etal2026}), of which \texttt{dustatt\_powerlaw}, \texttt{dustatt\_calzleit} and \texttt{dustatt\_modified\_CF00} fitted best and equally well, we adopt \texttt{dustatt\_powerlaw}. It uses one power law, $\alpha=-0.7$, for young ($<10$\,Myr) and old stars alike, but with separate normalisations: young stars are attenuated by both their birth clouds and the diffuse interstellar medium (ISM), and old stars only by the ISM. The V-band attenuation of the old stars is set to a fraction (\texttt{Av\_old\_factor}) of that of the young stars (\texttt{Av\_young}), and an optional UV bump at 217.5\,nm is included (amplitudes 0--3). The attenuations in the FUV, NUV and $u$, $g$ and $r$ bands are output by the module, and the $A_{\rm FUV}$ and $A_{\rm NUV}$ used in this paper are the total (young plus old) attenuations in the GALEX bands. For the power law with $\alpha=-0.7$, normalised at the V band (5500\,\AA), the young-star attenuation in the FUV is $A_{\rm FUV}\approx 2.4\,A_V$.

\subsection{Dust emission}
\label{sec:dust_module}

Interstellar dust emission has several components. Polycyclic aromatic hydrocarbons (PAHs) are nanometre-sized, transiently heated molecules that dominate the mid-IR emission near 8\,$\mu$m \citep{boquien2019cigale, roser2019pah}, whereas larger grains are in thermal equilibrium with the radiation field and produce the FIR emission. The dust distribution can be anisotropic, which affects the observed infrared emission \citep{zhao2024dust}.

We tested the empirical templates of \citet{dale2014two}, the models of \citet{draine2007infrared} (\texttt{dl2007}) and \citet{draine2014infrared} (\texttt{dl2014}), the analytic model of \citet{casey2012far} (\texttt{casey2012}) and the other dust-emission modules of \textsc{cigale} \citep{evans_etal2026}. In those tests, made on 50 of the 852 galaxies of the original sample (49 of which are in the final sample) with complete photometry from the FUV to 500\,$\mu$m, and with grids of similar size, the Draine \& Li models fitted best, with median reduced $\chi^2$ of 0.65 (\texttt{dl2007}) and 0.74 (\texttt{dl2014}), compared with 1.30 for \texttt{casey2012}; on these galaxies our final grid gives a median reduced $\chi^2$ 1.7 times that of the best \texttt{dl2007} grid \citep{evans_etal2026}. With the grids of the Thesis, \texttt{dl2014} also fitted the full sample better than \texttt{casey2012}.\footnote{For \texttt{dl2014} (6.5 million models), 233/798 (29 per cent) of the galaxies have a reduced $\chi^2<2.0$ and 480/798 (60 per cent) have a reduced $\chi^2<3.0$, compared with 146/798 (18 per cent) and 361/798 (45 per cent) for \texttt{casey2012} with the Thesis grid (9.07 million models). The two fits used the same photometry, with the faint 1.4\,GHz flux densities entered as detections. With the extended grid and the radio upper limits used here, \texttt{casey2012} has a median reduced $\chi^2$ of 1.33, with 662/798 (83 per cent) below 2.0 and 746/798 (93 per cent) below 3.0; the UV-to-submillimetre photometry is the same, but because of the different treatment of the radio values these runs are not comparable like for like with the Thesis fits.} A like-for-like comparison confirms this: a fit of the full sample with the radio upper limits used here and the best \texttt{dl2007} grid of the tests (15.8 million models) gives a median reduced $\chi^2$ of 0.85 for the \Nsamp\ galaxies, compared with 1.33 for our final grid, and a lower $\chi^2$ for 623 of them (78 per cent). We nevertheless use \texttt{casey2012}, because one aim of our analysis is to determine the dominant dust temperature, which the Draine \& Li models do not provide. Two other options in \textsc{cigale} v2022.1 give a dust temperature and fitted the test galaxies better than \texttt{casey2012}: the templates of \citet{schreiber2018egg} (\texttt{schreiber2016}), which are parametrised by a dust temperature and include PAH emission (median reduced $\chi^2$ of 0.97), and the \texttt{dale2014} templates with an added cold modified blackbody (0.85), for which only three temperatures could be included within the memory budget and most galaxies chose the highest \citep{evans_etal2026}. We prefer \texttt{casey2012} because its temperature is that of a modified blackbody with a free emissivity index, which can be compared directly with the temperatures of FIR-selected galaxies in the literature, and because it was the model used in the Thesis. The \texttt{schreiber2016} temperature is defined differently, as that of the template's dust continuum, but it ranks the galaxies in a similar way: on the 50 test galaxies it is lower than the \texttt{casey2012} temperature of our final grid by a median of 11.5\,K (interquartile range 7.5--15.9\,K), the two are correlated (Spearman rank coefficient 0.63), and the dust luminosities agree to a median of 0.05\,dex. The temperatures are observed-frame values (Section~\ref{sec:redshift}); their distribution is shown in Section~\ref{sec:timescales} and discussed in Section~\ref{sec:discussion}. The \texttt{casey2012} model is described by the dust temperature, the emissivity index $\beta$ and a mid-IR power-law slope $\alpha$. It combines a mid-IR power law, which approximates hot dust heated by AGN or clumpy starbursts, with a single-temperature modified blackbody that represents the reprocessed stellar emission in the FIR \citep{boquien2019cigale}. It is flexible and easy to use, but it does not model PAH emission, and since the infrared is computed from the energy balance it omits AGN heating \citep{boquien2019cigale}.

\subsection{Radio emission}
\label{sec:radio_module}

The \textsc{cigale} nebular module produces the thermal radio continuum but not the non-thermal synchrotron emission. The \texttt{radio} module instead uses the radio--IR correlation, assuming that the 21\,cm (1.4\,GHz) emission is mainly non-thermal. The infrared luminosity is used to predict the 21\,cm luminosity density, and a power-law radio spectrum (slope $\alpha_{\rm sf}=0.8$) is generated and scaled accordingly \citep{boquien2019cigale}. The infrared--radio relation is parametrised by
\begin{equation}
	q_{\rm IR} = \log_{10}\!\left(\frac{L_{\rm IR}}{3.75\times10^{12}\,\mathrm{W}}\right) - \log_{10}\!\left(\frac{L_{\rm 1.4\,GHz}}{\mathrm{W\,Hz^{-1}}}\right),
	\label{eq:qir}
\end{equation}
and we use the default value for star-forming galaxies, \texttt{qir\_sf}\,=\,2.58 \citep{helou1985thermal, yang2022fitting}. This is typical of the radio--IR correlation in the local Universe, but it also means that the model radio luminosity is determined by $L_{\rm IR}$. The observed 1.4\,GHz flux density is one of the 22 fitted bands, so a galaxy whose radio emission departs from $q_{\rm IR}=2.58$ is fitted less well in that band. For the 84 galaxies with a measured flux density the radio band acts as a constraint on $L_{\rm IR}$; for the 714 galaxies with upper limits it only excludes models in which the radio flux density predicted from $L_{\rm IR}$ exceeds the limit (Section~\ref{sec:data}). This has consequences for how the FIR--radio relation can be interpreted (Section~\ref{sec:radio}).

\subsection{AGN}
\label{sec:agn}

\textsc{cigale} offers several ways to include an AGN, from fast and crude (the AGN fraction, or the power-law slope in \texttt{casey2012}) to detailed (\texttt{fritz2006}, \citealt{fritz2006revisiting}; \texttt{skirtor2016}, \citealt{stalevski2012skirtor, stalevski2016dust}). AGN are thought to have a large effect on galaxy evolution \citep{boquien2019cigale}. An AGN component multiplies the size of the grid. A fit that can distinguish galaxies with and without an AGN needs at least two values of the AGN fraction (zero and one non-zero value), which doubles the number of models, and each further AGN parameter that is varied multiplies it again. Within the memory limit of the laptop (Section~\ref{sec:cigale}) this would require the rest of the grid to be halved, which would bring back the pile-up of best-fitting values at the edges of the grid. We therefore exclude from the sample the quasars and the galaxies whose mid-infrared emission is dominated by an AGN (Section~\ref{sec:selection}), and fit the remaining galaxies without an AGN component. The mid-infrared power law of \texttt{casey2012} (slope $\alpha$) can absorb some warm-dust emission from a weak AGN. The sample may still contain low-luminosity or composite AGN, and AGN that are bright only in the radio; their contributions to the UV, IR and radio emission are attributed to star formation in our fits, and we return to this in Section~\ref{sec:discussion}.

\subsection{Treatment of redshift}
\label{sec:redshift}

\textsc{cigale} groups the redshifts of the galaxies by rounding them to two decimal places, and runs the full model grid in each redshift bin \citep{boquien2019cigale}. The 852 galaxies of the Thesis fall in 77 bins, and the \Nsamp\ of this paper in 64, so the $1.63\times10^{7}$-model grid of this paper would become $1.63\times10^{7}\times64\approx1.05\times10^{9}$ models. At the time of the Thesis work no HPC facility was available to us, and there is none in Botswana. A first test on the mini PC, with the smaller grid of the Thesis ($9.07\times10^{6}$ models, $6.99\times10^{8}$ in 77 bins) and \textsc{cigale} v2022.1, using 8 cores and dividing the grid into 36 blocks, took 133.8\,h (5.6 days), about 140 times longer than the same grid at $z=0$ (0.96\,h). We found later that most of this time was lost: with $1.9\times10^{7}$ models per block the run used up to 27\,GB of the 32\,GB of RAM, and about half of the blocks took between 1.3 and 60 times as long as the others, probably because of memory pressure \citep{evans_memory2026}. After the analysis of this paper was complete, we fitted the \Nsamp\ galaxies at their true redshifts with our final grid and \textsc{cigale} v2025.1 on the same mini PC, dividing the grid into 105 blocks of about $10^{7}$ models. This fit took 12.2\,h on six cores, nine times as long as the same fit at $z=0$ with the same version (1.3\,h), and not 64 times, which suggests that most of the cost of each model is not repeated in each redshift bin. Its results are compared with those at $z=0$ below. A fit of this length is possible on a desktop computer, but it is exposed to power cuts, because \textsc{cigale} saves no results until it finishes \citep{evans_memory2026}, and on the evidence of the first test the true redshifts would have left no room for the repeated fits that the choice of modules and parameter ranges needed \citep{evans_etal2026}. We have since also obtained 24\,h of computing time on an HPC cluster, which we used for the checks below with the 50 test galaxies and for the fits of the full sample with \textsc{cigale} v2025.1 quoted in Sections~\ref{sec:results} and \ref{sec:discussion}. For the fits on which our results are based, we therefore set the redshift of every galaxy to zero in the \textsc{cigale} input, and scaled the quantities that depend on distance afterwards. Luminosities and SFRs scale as the square of the luminosity distance, and we calculated $D_{\rm L}$ for each galaxy from its catalogue redshift for a $\Lambda$CDM cosmology with $H_0=69.6\,\mathrm{km\,s^{-1}\,Mpc^{-1}}$, $\Omega_{\rm M}=0.286$ and $\Omega_\Lambda=0.714$ \citep{Bennett2014}, which is flat to within $10^{-4}$ in the curvature parameter. At $z=0$ \textsc{cigale} normalises each model to a luminosity distance of 10\,pc, so the luminosities, SFRs and stellar masses from the fits are multiplied by $(D_{\rm L}/10\,\mathrm{pc})^2$, which is between $2.5\times10^{13}$ and $2.1\times10^{17}$ for our galaxies.

The main consequence is that the fits contain no $k$-corrections: the observed-frame photometry is fitted as if it were in the rest frame. For the bulk of the sample (median $z=0.12$, third quartile $z=0.23$) the resulting shifts are modest, but they are not negligible, and they differ between bands. For example, at $z=0.25$ the GALEX FUV band samples the rest-frame wavelength $\approx 1220$\,\AA, and the $k$-correction in the SPIRE bands is strongly negative. The fitted dust temperature is also an observed-frame temperature, lower than the rest-frame value by a factor of about $(1+z)$. These effects increase with redshift, and we expect them to affect band-specific quantities (the FUV and NUV attenuations, individual band luminosities, the dust temperature) more than bolometric ones (the dust luminosity, the SFR).

We have measured the size of these effects with the 50 test galaxies of \citet{evans_etal2026}, which span $z=0.02$--0.47 (28 redshift bins). We fitted them with our final grid at $z=0$ and at their true redshifts, with \textsc{cigale} v2025.1 in both cases, on the HPC cluster (the fit at the true redshifts has $4.6\times10^{8}$ models), and compare the 45 galaxies for which the best fit at $z=0$ has no AGN component.\footnote{The $z=0$ fit included an optional \texttt{skirtor2016} component with an AGN fraction of 0 or 0.1 (Section~\ref{sec:discussion}); for these 45 galaxies the best fit is the same as with our final grid.} After the distance scaling, the dust luminosities agree to a median of $-0.02$\,dex (true-redshift minus $z=0$; Spearman coefficient 0.99) and the SFRs to $-0.02$\,dex (0.97). The stellar masses are lower at the true redshifts by a median of 0.11\,dex (interquartile range $-0.20$ to $-0.02$\,dex). The FUV attenuation is lower by a median of 0.12\,mag, but the difference grows with redshift: it is $-0.02$\,mag for $z<0.1$, $-0.11$\,mag for $0.1<z<0.2$ and $-0.85$\,mag for the ten galaxies at $z>0.2$. The dust temperature is higher at the true redshifts by a median of 5\,K, and the difference grows with redshift (Spearman coefficient 0.66); $T/(1+z)$ from the fits at the true redshifts agrees with the $z=0$ temperature to a median of 0.7\,K, as expected for an observed-frame temperature. At the true redshifts 11 of the 50 galaxies are at the upper edge of the temperature grid (50\,K), compared with six at $z=0$. We therefore repeated the fit at the true redshifts with the temperature grid shifted to 22.5--60\,K (no galaxy had a best-fitting temperature below 25\,K). Six of the 11 galaxies move to 55 or 60\,K and the other five stay at 50\,K, so that five galaxies (10 per cent) are still at the upper edge of the grid; the other 39 galaxies have the same best fit as before. For the six galaxies that move, $L_{\rm dust}$ and the SFR increase by medians of 0.01 and 0.06\,dex, and the medians for the sample as a whole do not change. The median reduced $\chi^2$ is slightly higher at the true redshifts (1.41 with the 50\,K grid and 1.38 with the 60\,K grid, compared with 1.30 at $z=0$), except for the galaxies at $z>0.2$, which are fitted better. A test with the \texttt{dl2007} grid, in which both fits were made with v2022.1, gives similar results (dust luminosity $+0.01$\,dex, SFR and stellar mass $-0.05$\,dex, $A_{\rm FUV}$ $-0.09$\,mag, and $-0.49$\,mag at $z>0.2$). The $z=0$ approximation therefore has little effect on the dust luminosities and SFRs on which most of our results are based, but it lowers the temperatures by a factor of about $(1+z)$, biases the stellar masses high by about 0.1\,dex, and overestimates the FUV attenuation of the galaxies at $z>0.2$ (30 per cent of our sample) by up to about 1\,mag.

The fit of the whole sample at the true redshifts confirms these results. We compare it with a fit at $z=0$ made with the same grid and the same version of \textsc{cigale} (v2025.1), so that only the redshift differs; in the fit at the true redshifts \textsc{cigale} computes the luminosity distances for the cosmology of \citet{planck2020}, and we rescaled its luminosities, SFRs and stellar masses to our cosmology (a median change of $-0.02$\,dex). The dust luminosities agree to a median of $-0.01$\,dex (true-redshift minus $z=0$; interquartile range $-0.05$ to $+0.03$\,dex) and the SFRs to $0.00$\,dex ($-0.07$ to $+0.09$\,dex); for the 243 galaxies at $z>0.2$ the SFRs are higher at the true redshifts by a median of 0.10\,dex. The stellar masses are lower at the true redshifts by a median of 0.14\,dex (interquartile range $-0.28$ to $-0.04$\,dex): by 0.05, 0.16 and 0.33\,dex for the galaxies at $z<0.1$, $0.1<z<0.2$ and $z>0.2$. At the true redshifts no model can be older than the Universe at the redshift of the galaxy, whereas at $z=0$ the best-fitting ages of 120 galaxies exceed it, which may contribute to this difference. The FUV attenuation is lower by a median of 0.17\,mag, and by 0.04, 0.15 and 1.13\,mag in the three redshift ranges. The dust temperature is higher by a median of 5\,K (2.5, 5 and 12.5\,K in the three ranges; Spearman coefficient between the difference and the redshift 0.67), and $T/(1+z)$ from the fit at the true redshifts agrees with the $z=0$ temperature to a median of 0.8\,K. At the true redshifts 149 galaxies (19 per cent) are at the upper edge of the temperature grid (50\,K), compared with 36 at $z=0$ with the same version, so a fit at the true redshifts needs a temperature grid extending above 50\,K; the test galaxies suggest that this would change the dust luminosities and SFRs very little. The median reduced $\chi^2$ is 1.28, compared with 1.27 at $z=0$: it is higher at the true redshifts for the galaxies at $z<0.1$ (1.37 compared with 1.19) and $0.1<z<0.2$ (1.41 and 1.23), and lower for those at $z>0.2$ (1.02 and 1.36). For the sample as a whole, the $z=0$ approximation therefore changes the dust luminosities and SFRs by about 0.01\,dex in the median, but it lowers the dust temperatures by a factor of about $(1+z)$, biases the stellar masses high by a median of 0.14\,dex (0.33\,dex at $z>0.2$), and overestimates the FUV attenuation of the galaxies at $z>0.2$ by a median of 1.1\,mag.

\section{Star formation rate indicators}
\label{sec:indicators}

We consider two kinds of SFR. The first are the SFRs of the best-fitting \textsc{cigale} models, which are constrained by the whole SED and are averaged over three timescales: the instantaneous SFR, and the SFRs averaged over the last 10 and 100\,Myr, which we denote $\sfr{inst}$, $\sfr{10}$ and $\sfr{100}$. The second are the single-band indicators,
\begin{align}
	\sfr{UV} &= C_{\rm FUV}\, L_{\rm FUV}, \label{eq:sfr_uv}\\
	\sfr{FIR} &= C_{\rm IR}\, L_{\rm IR}, \label{eq:sfr_fir}\\
	\sfr{Radio} &= C_{\rm 1.4}\, L_{\rm 1.4\,GHz}, \label{eq:sfr_radio}
\end{align}
where $L_{\rm FUV}$ is the observed (uncorrected) FUV luminosity $\nu L_\nu$, $L_{\rm IR}$ is the dust (total infrared) luminosity, and $L_{\rm 1.4\,GHz}$ is the 1.4\,GHz luminosity density. An attenuation-corrected UV SFR is obtained by multiplying $L_{\rm FUV}$ by $10^{0.4A_{\rm FUV}}$. For the UV and FIR we use the calibrations of \citet{kennicutt1998star}, $C_{\rm FUV}=1.4\times10^{-28}\,\Msun\,\mathrm{yr^{-1}}$ per $\mathrm{erg\,s^{-1}\,Hz^{-1}}$ (with the FUV luminosity density $L_\nu$) and $C_{\rm IR}=4.5\times10^{-44}\,\Msun\,\mathrm{yr^{-1}}$ per $\mathrm{erg\,s^{-1}}$. These are for a Salpeter IMF ($0.1$--$100\,\Msun$), the IMF used in our fits, so they need no conversion. We do not compute $\sfr{Radio}$ from the fitted radio luminosities, because in the models these are determined by $L_{\rm IR}$ through Eq.~(\ref{eq:qir}), so that such an SFR would be proportional to $\sfr{FIR}$; the radio emission is discussed in Section~\ref{sec:radio}. The \citet{kennicutt1998star} UV calibration assumes a flat $L_\nu$ between 1500 and 2800\,\AA, and the GALEX FUV band (effective wavelength 1528\,\AA) lies at the short-wavelength end of this range. Expressed per unit $\nu L_\nu$ at 1528\,\AA, it gives $7.1\times10^{-44}\,\Msun\,\mathrm{yr^{-1}}$ per $\mathrm{erg\,s^{-1}}$, within 0.03\,dex of the FUV calibration of \citet{KennicuttEvans2012} ($10^{-43.35}$ for a Kroupa IMF) converted to the Salpeter IMF with a factor of 1.5, so the choice of calibration does not affect our results. We compare these with each other and with the SED-based SFRs in Section~\ref{sec:comparison}.

\section{Results}
\label{sec:results}

All \textsc{cigale} fits succeeded. For the \Nsamp\ galaxies the median reduced $\chi^2$ is 1.33, and it is below 2 for 662 galaxies (83 per cent) and below 3 for 746 (93 per cent). All the results are based on the \Nsamp\ galaxies, and fractions are fractions of \Nsamp.

\subsection{FUV attenuation and FIR luminosity}
\label{sec:fuv_fir}

Dust absorbs the light of stars and AGN, and re-radiates the energy in the FIR and sub-millimetre, so that dust is the main source of emission at those wavelengths. It is difficult to determine how much starlight is attenuated, and how this varies with wavelength and environment \citep{viaene2016bolometric}. \citet{hao2011dust} calibrated the infrared excess, $\log\left[L(\mathrm{TIR})/L(\mathrm{FUV})_{\rm obs}\right]$ (IRX), and the FUV--NUV colour as indicators of FUV attenuation. With an rms scatter of $\pm0.09$\,dex ($\pm0.23$\,mag), the IRX-corrected FUV luminosities from their calibration are linearly related to H$\alpha$ luminosities corrected using the Balmer decrement. \citet{Casey_2014} reached a similar conclusion in a study of attenuation in dusty star-forming galaxies selected at far-infrared wavelengths. They also found that the SFR and the UV attenuation are strongly correlated, and that galaxies with $\mathrm{SFR}>50\,\Msun\,\mathrm{yr^{-1}}$ deviate from the IRX--$\beta$ relation, indicating that higher FIR luminosities are associated with greater UV absorption.

FIR and FUV are thus complementary, combining the obscured and unobscured components of star formation \citep{kennicutt1998star}. The FIR luminosity measures the obscured SFR, since dust grains absorb a large fraction of the UV and optical light of young stars and re-radiate it in the FIR \citep{kennicutt1998star}, while the FUV attenuation measures the fraction of UV light absorbed by dust, and is the quantity needed to correct UV-derived SFRs.

\begin{figure*}
	\includegraphics[width=0.9\textwidth]{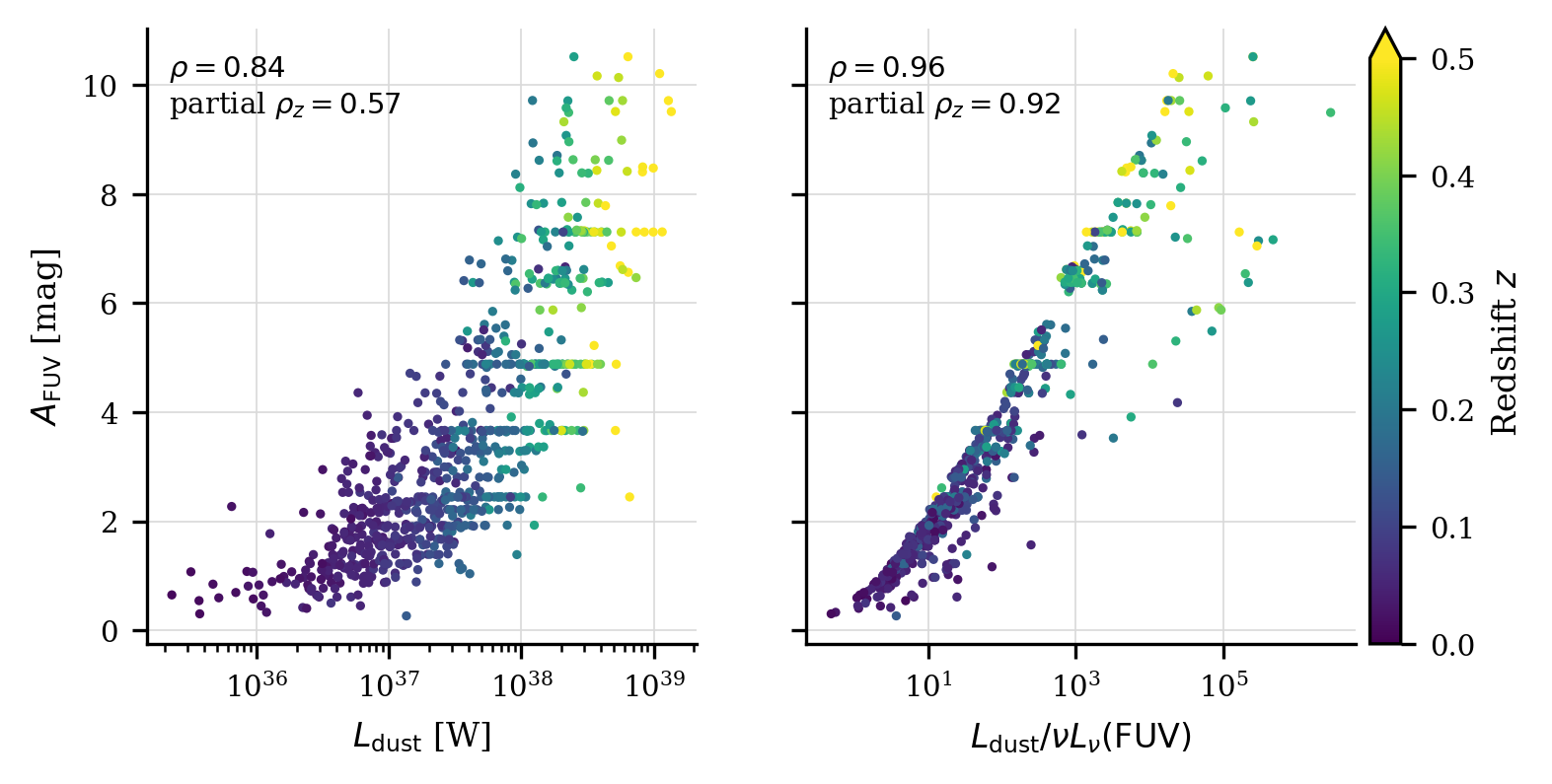}
	\caption{FUV attenuation as a function of the dust luminosity (left) and of the infrared excess $L_{\rm dust}/\nu L_\nu(\mathrm{FUV})$ (right), colour-coded by redshift. The Spearman coefficient is given in each panel; the partial coefficient controls for redshift. Unlike the luminosity, the infrared excess does not depend on distance.}
	\label{fig:afuv_redshift}
	\par\smallskip{\small Alt text: Two scatter plots of FUV attenuation, from 0 to 10.5\,mag, against the dust luminosity (left) and against the ratio of dust to FUV luminosity (right), with points coloured by redshift from 0 to 0.5. The left panel shows a broad rising trend, with the higher-redshift galaxies at the highest luminosities; the right panel shows a much narrower rising relation.}
\end{figure*}

The left panel of Fig.~\ref{fig:afuv_redshift} shows the FUV attenuation as a function of the dust (FIR) luminosity. There is a strong positive Spearman correlation ($\rho=0.84$), so the FUV attenuation rises with FIR luminosity. This agrees with the findings of \citet{10.1111/j.1365-2966.2005.09131.x} and \citet{hao2011dust}, and with the expectation that the FIR emission is mainly from dust heated by the UV and optical photons of young massive stars \citep{calzetti2000dust}. Dust is more prevalent in systems with high star formation activity, so these galaxies have both higher FIR luminosities and greater FUV attenuation. Some galaxies deviate from the trend, which may reflect differences in dust grain composition, in the spatial distribution of the dust, or in the SFH \citep{calzetti2000dust}.

This correlation should not be interpreted as independent evidence that dust drives the trend. In the energy-balance modelling, the dust luminosity is set by the UV/optical energy absorbed, and the attenuation is the model quantity that determines that energy, so that a positive relation between the two is partly built into the fit. The figure shows that the fits are self-consistent, but it does not test the physical picture.

A second, more important, qualification is that both quantities depend strongly on redshift. In a flux-limited sample the dust luminosity increases with redshift (Spearman $\rho=0.93$ between $L_{\rm dust}$ and $z$), and so does the attenuation ($\rho=0.75$; the median $A_{\rm FUV}$ rises from 1.7\,mag for $z<0.08$ to 6.5\,mag for $z>0.3$). At fixed redshift the partial rank correlation between $A_{\rm FUV}$ and $L_{\rm dust}$ falls to 0.57, and within five redshift bins with equal numbers of galaxies it is 0.49--0.65. We cannot tell from these fits whether the rise of $A_{\rm FUV}$ with redshift is physical (more luminous, dustier galaxies at greater distances) or partly an artefact of fitting observed-frame fluxes at $z=0$, so that the FUV band samples ever shorter rest-frame wavelengths (Section~\ref{sec:redshift}); the fits at the true redshifts of the test galaxies suggest that the latter accounts for about 1\,mag of the attenuation at $z>0.2$. A relation that does not depend on distance is that between $A_{\rm FUV}$ and the infrared excess $\mathrm{IRX}=L_{\rm dust}/\nu L_\nu(\mathrm{FUV})$, for which $\rho=0.96$ for the whole sample and 0.87--0.96 within the redshift bins (Fig.~\ref{fig:afuv_redshift}, right); as discussed above, this is also partly built into the models.

The highest fitted dust luminosity in the sample is that of HATLAS~J084734.8+015637 ($z=0.63$; $L_{\rm dust}=1.4\times10^{39}$\,W, $\approx3.6\times10^{12}\,L_\odot$), and the largest SFR is that of HATLAS~J091825.8+014720 ($z=0.55$; $350\,\Msun\,\mathrm{yr^{-1}}$ averaged over 100\,Myr). All seven sources of the Thesis sample at $z>1$, which had fitted luminosities of up to $1.7\times10^{40}$\,W, were quasars and are not in the final sample.

\subsection{FUV and NUV emission and attenuation in the FIR bands}
\label{sec:uv_fir_attenuation}

\begin{figure*}
	\includegraphics[width=\textwidth]{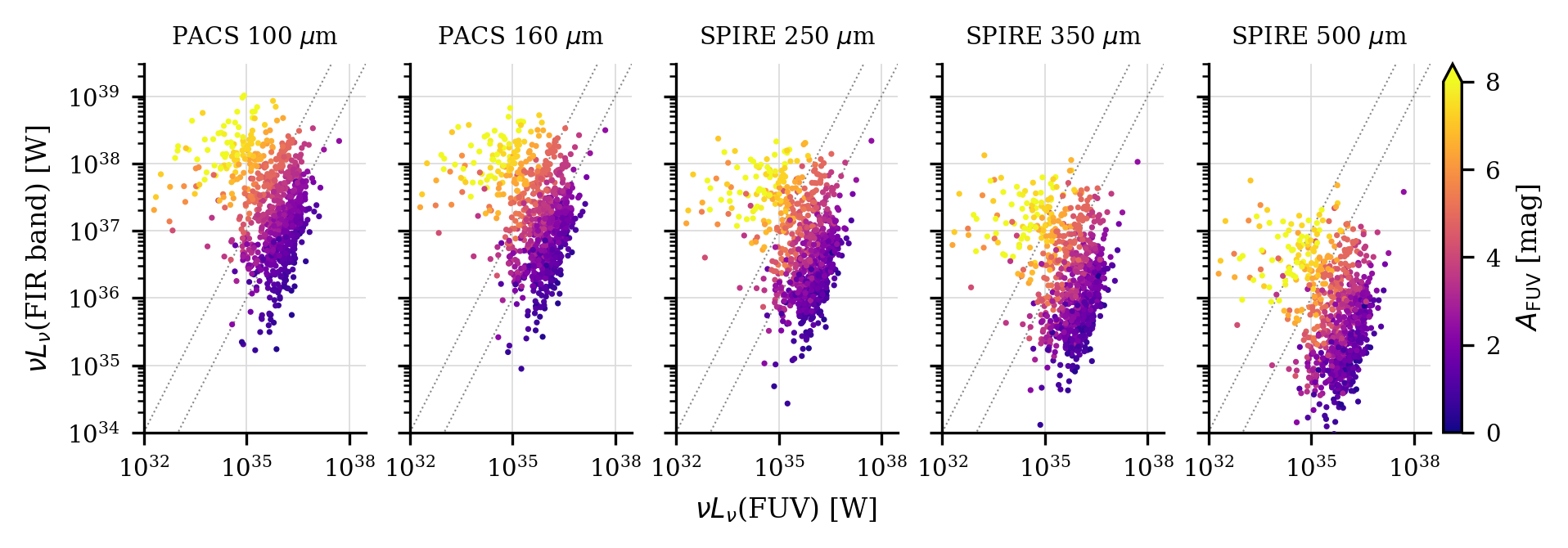}
	\caption{FIR band luminosity against FUV luminosity for the five individual FIR bands (PACS 100 and 160\,$\mu$m; SPIRE 250, 350 and 500\,$\mu$m). The colour coding shows the FUV attenuation, as indicated by the colour bar to the right of the panels; the arrow at its top means that galaxies with $A_{\rm FUV}>8$\,mag are shown in the colour for 8\,mag. The dotted grey lines mark $\nu L_\nu(\mathrm{FIR\ band})=10$ and $100\,\nu L_\nu(\mathrm{FUV})$, lines of constant infrared-to-UV ratio.}
	\label{fig:FUV vs FIR}
	\par\smallskip{\small Alt text: Five logarithmic scatter panels, one for each \textit{Herschel} band from 100 to 500\,$\mu$m, of the FIR band luminosity against the FUV luminosity, coloured by FUV attenuation from 0 (purple) to 8\,mag (yellow). Galaxies with low attenuation lie to the lower right and those with high attenuation to the upper left, above two dotted diagonal lines of constant luminosity ratio.}
\end{figure*}

\begin{figure*}
	\includegraphics[width=\textwidth]{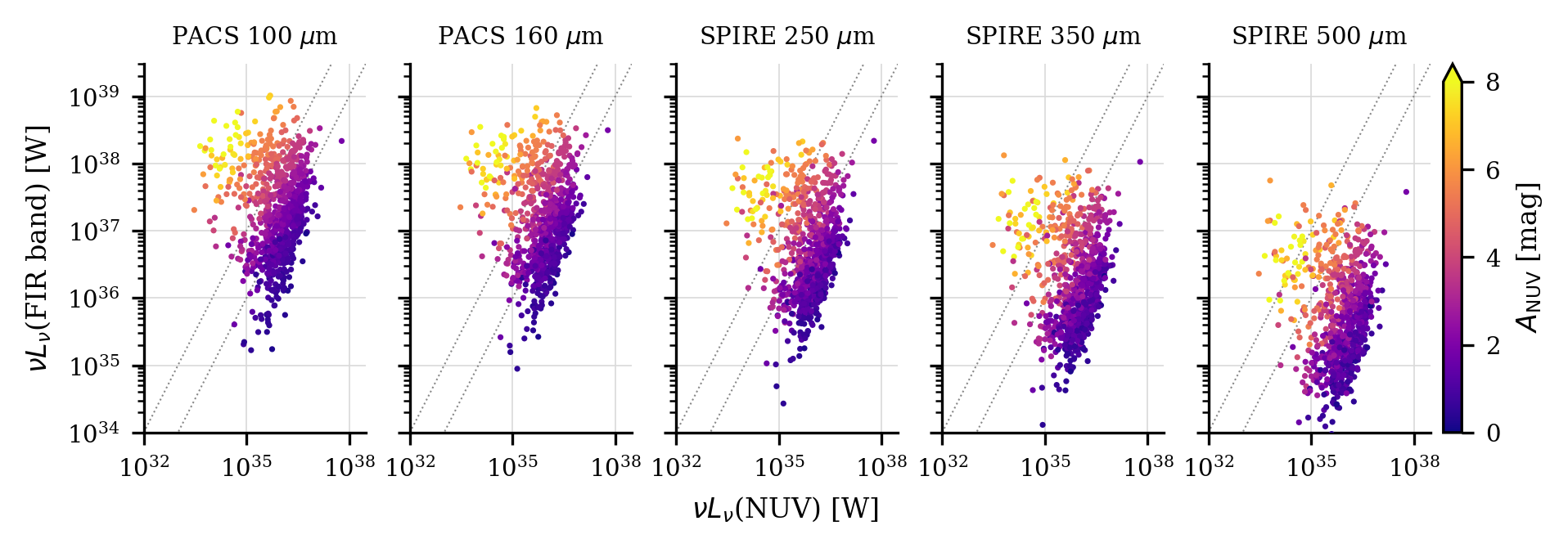}
	\caption{As Fig.~\ref{fig:FUV vs FIR}, but for the NUV luminosity. The colour coding shows the NUV attenuation, as indicated by the colour bar to the right of the panels (galaxies with $A_{\rm NUV}>8$\,mag are shown in the colour for 8\,mag). The dotted grey lines mark $\nu L_\nu(\mathrm{FIR\ band})=10$ and $100\,\nu L_\nu(\mathrm{NUV})$, lines of constant infrared-to-UV ratio.}
	\label{fig:NUV vs FIR}
	\par\smallskip{\small Alt text: Five logarithmic scatter panels, one for each \textit{Herschel} band from 100 to 500\,$\mu$m, of the FIR band luminosity against the NUV luminosity, coloured by NUV attenuation. As in the previous figure, galaxies with low attenuation lie to the lower right and those with high attenuation to the upper left, above two dotted diagonal lines of constant luminosity ratio.}
\end{figure*}

Figures~\ref{fig:FUV vs FIR} and~\ref{fig:NUV vs FIR} show, for each of the five \textit{Herschel} bands (PACS 100 and 160\,$\mu$m; SPIRE 250, 350 and 500\,$\mu$m), the FIR band luminosity plotted against the GALEX FUV and NUV luminosity respectively. In both figures the points are colour-coded by the attenuation ($A_{\rm FUV}$ or $A_{\rm NUV}$) from the \textsc{cigale} fits. The attenuation is carried by the colour scale and is not plotted on an axis, so the information it provides is in how the colour varies across the luminosity--luminosity plane.

\textit{Interpretation as an infrared-excess diagram.} On logarithmic axes, lines of constant $L_{\rm FIR}/L_{\rm UV}$ are diagonals of unit slope, such as the dotted lines in Figs~\ref{fig:FUV vs FIR} and~\ref{fig:NUV vs FIR}, and this ratio is, to within a bolometric correction, the infrared excess, $\mathrm{IRX}=L_{\rm IR}/L_{\rm UV}$ \citep{Meurer1999, Buat2005}. The figures are therefore the luminosity--luminosity counterpart of the IRX--attenuation relation. In all ten panels the galaxies form a broad cloud with only a mild overall positive slope, and the colour varies systematically \emph{across} the cloud and not along it. The most attenuated galaxies (yellow) lie in the upper left, where the FIR is bright relative to the UV, and those with low attenuation (purple) lie in the lower right, where the UV is bright relative to the FIR. The high-attenuation galaxies form a fairly narrow diagonal band, which is the locus of approximately constant, large IRX. This is expected if the FIR emission is dust-reprocessed UV light, and it agrees with the IRX--attenuation calibrations of \citet{10.1111/j.1365-2966.2005.09131.x} and \citet{hao2011dust} and with the findings of \citet{Casey_2014} for dusty star-forming galaxies. It also shows that the energy-balance fits assign larger attenuations to the galaxies that the data identify as having the largest FIR/UV ratios.

This bears directly on the aim of the paper. An uncorrected UV SFR scales as $L_{\rm UV}$, while the FIR SFR scales as $L_{\rm FIR}$, so the vertical offset of a galaxy above a line of constant IRX measures the factor by which $\sfr{UV}$ underestimates the obscured star formation. In magnitudes, $A_{\rm FUV}=2$\,mag corresponds to a loss of FUV flux by a factor of $10^{0.4\times2}\approx6$, and $A_{\rm FUV}=5$\,mag to a factor of $\approx100$. The galaxies in the upper left of the panels are therefore those for which the uncorrected UV is least reliable, which is the visual basis for the result in Section~\ref{sec:comparison} that the discrepancies grow for $A_{\rm FUV}>2$\,mag. Conversely, the galaxies in the lower right are those in which the UV carries most of the star formation signal and an uncorrected UV SFR is least biased.

\textit{Comparison of the FUV and NUV diagrams.} Fig.~\ref{fig:NUV vs FIR} shows the same trend as Fig.~\ref{fig:FUV vs FIR}, with somewhat lower attenuations. This is expected from the adopted attenuation curve. For the power law of Eq.~(\ref{eq:powerlaw}) with $\alpha=-0.7$ and the effective GALEX wavelengths of 1528\,\AA\ (FUV) and 2271\,\AA\ (NUV),
\begin{equation}
	\frac{A_{\rm NUV}}{A_{\rm FUV}} \approx \left(\frac{2271}{1528}\right)^{-0.7} \approx 0.76,
\end{equation}
so a galaxy with $A_{\rm FUV}=4$\,mag should have $A_{\rm NUV}\approx3$\,mag. This ratio holds for each stellar component separately; for the combined young plus old population the ratio is only approximately constant, because the two components contribute differently in the two bands. The offset between the two colour scales is therefore a consistency check on the fits, and not an independent result. The NUV is the band most sensitive to the UV bump, which lies at 2175\,\AA\ \citep{Noll2009} within the NUV passband, and our grid allows bump amplitudes between 0 and 3 (Table~\ref{tab:cigale_parameters}). A departure of the NUV panels from the scaling above would therefore most plausibly reflect the bump and, more generally, the form of the attenuation law \citep[see][]{SalimNarayanan2020}, and not differences in the stellar populations.

\textit{Dependence on FIR band.} The five panels of each figure are not five independent tests. In the \texttt{casey2012} model the FIR emission of each galaxy is described by a single modified blackbody, so the five bands are strongly correlated and similar-looking panels are expected. What the comparison does show is the dependence on dust temperature. For the dust temperatures of our grid (17.5--50\,K), PACS 100\,$\mu$m lies close to, or on the short-wavelength side of, the peak of the dust emission, and is therefore the band most affected by the dust temperature. SPIRE 500\,$\mu$m lies on the Rayleigh--Jeans tail and traces the colder dust and the dust mass more closely \citep{casey2012far}. If the colour gradient is cleaner in the longer-wavelength bands, this indicates sensitivity to the dust temperature; if it is equally clear in all five, the result does not depend on the choice of FIR band. In our figures the colour gradient across the cloud is equally clear in all five bands, so the result does not depend on the choice of FIR band.

\textit{Caveats.} Several caveats apply to Figs.~\ref{fig:FUV vs FIR} and~\ref{fig:NUV vs FIR}.
\begin{enumerate}[label=(\roman*)]
\item \emph{Partial circularity.} As in Section~\ref{sec:fuv_fir}, a correlation between the FIR luminosity and attenuation is partly built in by the energy-balance modelling \citep{boquien2019cigale}. The figures show that the fits are self-consistent, but they are not independent evidence that dust alone drives the trends.
\item \emph{Shared distance dependence.} Both axes are luminosities computed with the same luminosity distance, so a positive correlation between them is induced, even in the absence of any physical link, in a sample extending over a range of redshifts (the Spearman coefficient between $L_{\rm dust}$ and $z$ is 0.94). The constant-IRX diagonals are unaffected, since the distance cancels in the ratio, but the overall slope of the cloud should not be interpreted physically.
\item \emph{Selection effects.} The sample is selected from SPIRE detections \citep{valiante2016herschel}, which favours FIR-bright, dusty galaxies. The sharp lower boundary of the cloud at low FIR luminosity is likely to be a flux limit and not a physical boundary, and the scarcity of galaxies with low attenuation is probably in part a selection effect. We therefore make no statement about the intrinsic frequency of high attenuation.
\item \emph{Extreme attenuations.} The largest attenuations on the colour scales are poorly constrained. For 74 galaxies $A_{\rm FUV}>7.2$\,mag, and for five it is above 10\,mag (up to 10.5\,mag), so that the observed FUV flux is reduced by a factor of $10^{3}$ or more, and the FUV band, where it is measured, carries almost no information on the attenuation. The best-fitting model has the largest \texttt{Av\_young} in the grid (8\,mag) for 1.4 per cent of the galaxies (11), and one of the five with $A_{\rm FUV}>10$\,mag is among them, so the attenuations at the top of the colour scale should be read as `large' and not as measurements. The best-fitting \texttt{age\_main} is at the upper edge of the grid (12\,Gyr) for 27 per cent of the galaxies.
\item \emph{Fixed redshift and outliers.} As discussed in Section~\ref{sec:redshift}, no $k$-corrections are applied. This is a good approximation for the bulk of the sample, but less so for the SPIRE bands and for the galaxies at $z>0.2$, whose FUV attenuations are overestimated by a median of about 1\,mag (Section~\ref{sec:redshift}). The UV-brightest sources of the Thesis sample ($\nu L_\nu(\mathrm{FUV})>10^{38}$\,W), which had low fitted attenuations and were probably UV-bright AGN, were all quasars and are not in the final sample.
\item \emph{Degeneracies.} Our SFHs allow ages up to 12\,Gyr (the table in Appendix~\ref{app:cigale}). In galaxies with low specific SFR, FUV emission from older stars is degenerate with attenuation, so the attenuations of the UV-bright (purple) galaxies are less well constrained than the colour scale suggests.
\end{enumerate}
In summary, Figs.~\ref{fig:FUV vs FIR} and~\ref{fig:NUV vs FIR} show that in our sample galaxies with large FIR/UV ratios are those with large fitted attenuations in both UV bands, as expected for dust-reprocessed star formation, and they illustrate qualitatively why uncorrected UV SFRs underestimate star formation in dusty systems. A quantitative test requires $\sfr{FIR}/\sfr{UV}$ (or $\log\mathrm{IRX}$) to be plotted directly against $A_{\rm FUV}$, which we do in Section~\ref{sec:comparison}.

\subsection{FIR luminosity and the radio continuum}
\label{sec:radio}

The FIR--radio correlation (FIRC) has minimal scatter \citep{read2018far}. Several processes contribute to it, including the efficient cooling of cosmic-ray electrons (calorimetry) in starbursts \citep{lacki2010physics}. Over a very broad range of SFRs, from normal spirals like the Milky Way to the most powerful starbursts, the FIR and radio luminosities of star-forming galaxies are linearly correlated \citep{lacki2010physics}. Since star formation is the main source of FIR emission, the correlation suggests that the radio power is also tied to star formation, and it has been used to establish a radio SFR calibration \citep{molnar2021non}. The FIRC is reported to be unchanged when the dust temperature, redshift and stellar mass vary \citep{read2018far}. The radio synchrotron continuum has three main advantages over other SFR tracers: it is not attenuated by dust, so it needs no correction; radio interferometers give high angular resolution; and next-generation telescopes have exceptional sensitivity and survey speed \citep{molnar2021non}.

\begin{figure}
	\includegraphics[width=\columnwidth]{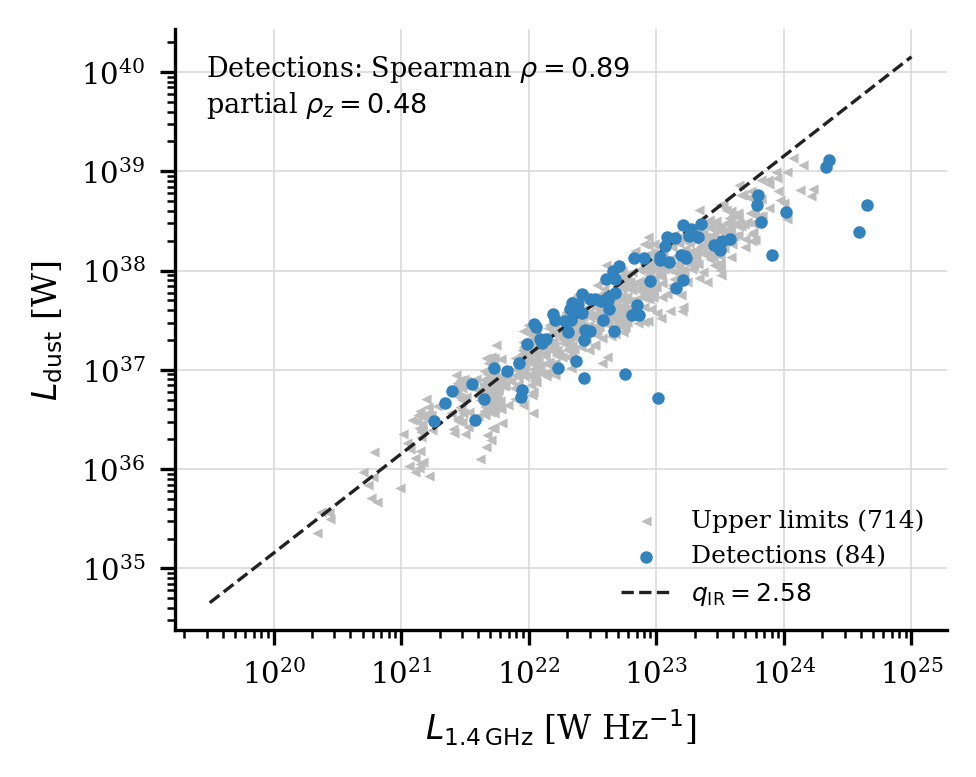}
	\caption{Dust luminosity of the best-fitting model against the observed 1.4\,GHz luminosity (from the FIRST flux density, $k$-corrected for a spectral index of 0.8) for the \Nsamp\ galaxies. Blue points: the 84 detections; grey triangles: the 714 upper limits, plotted at the luminosity of the 0.75\,mJy limit. The dashed line is $q_{\rm IR}=2.58$, the value assumed in the models. The Spearman and partial (at fixed redshift) coefficients are for the detections.}
	\label{fig:FIR_vs_Radio_Correlation.png}
	\par\smallskip{\small Alt text: Logarithmic scatter plot of the dust luminosity, from $10^{35}$ to $10^{39.5}$\,W, against the 1.4\,GHz luminosity, from $10^{20.5}$ to $10^{24.5}$\,W\,Hz$^{-1}$. Blue points for detections and grey arrows for upper limits follow a tight band along a dashed diagonal line for $q_{\rm IR}=2.58$, with a few detections well below the line at high radio luminosity.}
\end{figure}

Fig.~\ref{fig:FIR_vs_Radio_Correlation.png} shows the dust luminosity of the best-fitting model against the observed 1.4\,GHz luminosity, computed from the FIRST flux density with a $k$-correction for a spectral index of 0.8, for the \Nsamp\ galaxies; for the 714 upper limits the radio luminosity is that of the 0.75\,mJy limit. For the 84 detections the relation is close to linear on logarithmic axes, in agreement with \citet{Yun_2001} and \citet{magnelli2015far}, who found a similar relation for star-forming galaxies. There is some evidence of non-linearity in more recent work \citep{cox1988correlation, molnar2021non}.

Two points need to be made about this result. First, the observed 1.4\,GHz flux density is one of the fitted bands, so the figure is a test of the FIRC for the galaxies with measured flux densities, although the fitted $L_{\rm IR}$ is itself partly constrained by that flux density through the fixed $q_{\rm IR}$ (Section~\ref{sec:radio_module}). The model radio luminosities could not be used for such a test: in the \textsc{cigale} results they have $q_{\rm IR}=2.50$--2.61 (median 2.59) for the \Nsamp\ galaxies, so their correlation with $L_{\rm dust}$ is built in by construction through Eq.~(\ref{eq:qir}). Second, both luminosities scale with the square of the distance, and since $L_{\rm dust}$ is strongly correlated with redshift in this flux-limited sample, the redshift dependence will dominate any luminosity--luminosity correlation, so that the plot is not a good test of the FIRC, whereas $q_{\rm IR}$ is independent of distance.

The informative quantity is the observed $q_{\rm IR}$ (Eq.~\ref{eq:qir}) of each galaxy compared with the assumed value of 2.58. It can be computed only for the 84 galaxies with a measured flux density; for the 714 upper limits it gives upper limits on the radio luminosity and hence lower limits on $q_{\rm IR}$. We computed it from the 1.4\,GHz flux densities in the input photometry, with $L_{\rm IR}$ taken to be the fitted $L_{\rm dust}$, and a $k$-correction to the rest frame for a spectral index of 0.8. For the 84 galaxies the median is $q_{\rm IR}=2.51$ (mean 2.42, 16--84 per cent range 2.15--2.72, standard deviation 0.34\,dex), about 0.07\,dex below the value of 2.58 assumed in the models, and 71 per cent of them are within 0.3\,dex of 2.58. The observed radio and dust luminosities are strongly correlated (Spearman $\rho=0.89$, and a Pearson coefficient of 0.88 on logarithmic axes), but much of this is the common dependence on distance: at fixed redshift the partial coefficient is 0.48. The observed $q_{\rm IR}$ decreases with redshift ($\rho=-0.41$) and weakly with $L_{\rm dust}$ ($\rho=-0.17$), but this is largely a selection effect of the flux-density limit (see below). Because the fits assume $q_{\rm IR}=2.58$, the best-fitting models underestimate the observed 1.4\,GHz flux density of these galaxies by a median factor of 0.81 (the model flux density is within a factor of 2 of the observed one for 69 per cent), and they are fitted less well than the rest of the sample (median reduced $\chi^2$ of 1.71, compared with 1.28 for the galaxies with upper limits). The observed 100\,$\mu$m flux density agrees with the model (median ratio of model to observed 1.01). For the 714 upper limits, the model flux density is below the limit of 0.75\,mJy for 81 per cent (median 0.51\,mJy), as expected if their radio emission follows the FIRC with $q_{\rm IR}\approx2.6$, and the remaining 19 per cent exceed it and are penalised in the fit. The fixed $q_{\rm IR}$ is therefore a source of tension only for the radio-brightest galaxies, which may host radio emission from AGN (Section~\ref{sec:discussion}). The WISE criterion does not remove radio AGN: seven of the 84 detections have an observed $q_{\rm IR}$ more than 0.5\,dex below 2.58 (values of 1.13--2.00), and are probable radio AGN. Without them the median $q_{\rm IR}$ of the detections is 2.54, the standard deviation is 0.22\,dex, and 78 per cent are within 0.3\,dex of 2.58.

\begin{figure}
	\includegraphics[width=\columnwidth]{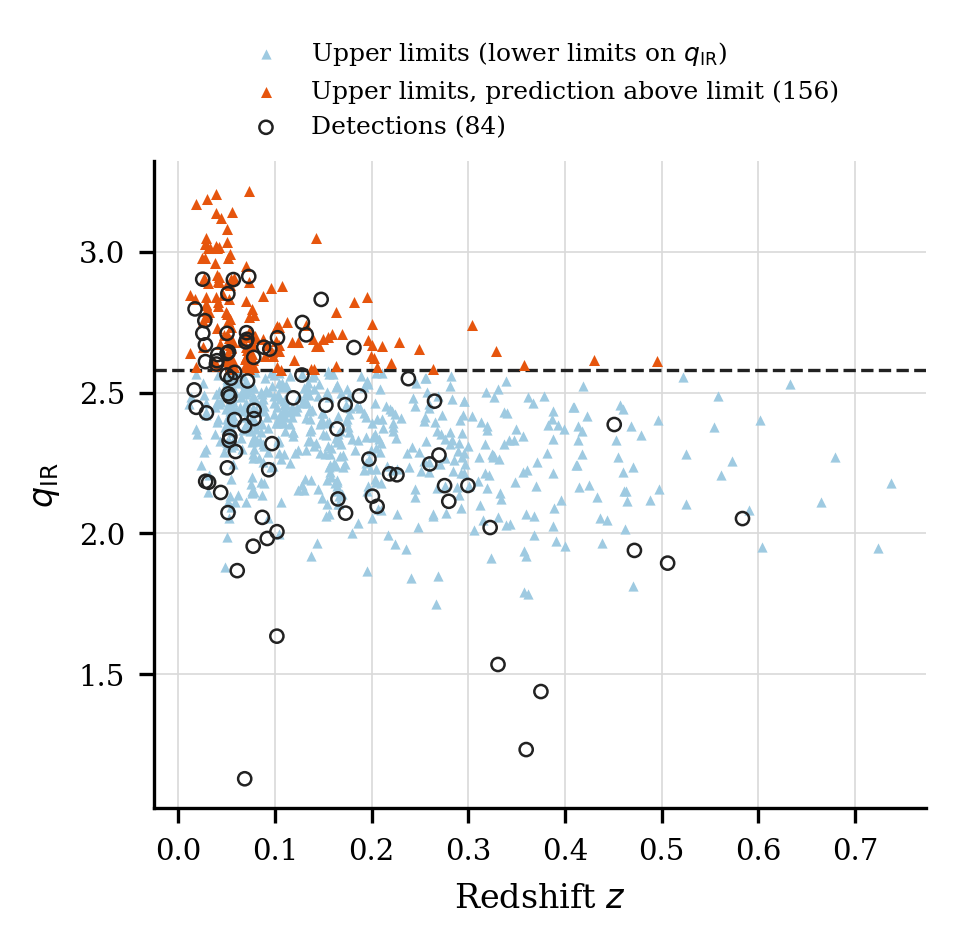}
	\caption{The infrared--radio ratio $q_{\rm IR}$ against redshift, for the fit without the radio band. Circles: the 84 galaxies with FIRST detections ($q_{\rm IR}$ from the observed flux density). Triangles: lower limits on $q_{\rm IR}$ for the 714 galaxies with upper limits (from the 0.75\,mJy limit), in orange where the predicted flux density exceeds the limit (156 galaxies). The dashed line is the value assumed in the models, $q_{\rm IR}=2.58$. The lower edge of the distribution of the triangles rises towards low redshift because a fixed flux-density limit is deeper in luminosity for nearby galaxies.}
	\label{fig:qir_redshift}
	\par\smallskip{\small Alt text: Scatter plot of $q_{\rm IR}$, from 1.1 to 3.3, against redshift from 0 to 0.75. Open circles for the 84 radio detections lie mostly between 2.0 and 2.9, around a horizontal dashed line at 2.58; small blue triangles for the upper limits lie mostly between 1.8 and 2.6, and orange triangles above 2.58 are concentrated below $z=0.2$.}
\end{figure}

The treatment of the faint values has a large effect on the fits. The two tests described next were made with the intermediate grid of 12.9 million models (Section~\ref{sec:cigale}), for which the fit with upper limits has a median reduced $\chi^2$ of 1.43 for the \Nsamp\ galaxies. In a test fit with that grid in which all the faint values were entered as detections with a 15 per cent uncertainty, the median reduced $\chi^2$ was 2.65, compared with 1.43 with upper limits, 73 per cent of the galaxies were best fitted by a temperature of 17.5\,K or lower, compared with 4 per cent, and the dust luminosities were lower by a median factor of 2.1. Noise-level radio values combined with the fixed $q_{\rm IR}$ therefore force the fit towards low $L_{\rm IR}$ and low dust temperatures, in conflict with the many FIR bands, and the use of upper limits is essential. To see how much the radio band affects the other results, and to obtain a test of the FIR--radio relation that does not depend on the fit, we repeated the fit with the intermediate grid and the same photometry but without the 1.4\,GHz band and without the radio module. The median reduced $\chi^2$ is 1.42 (compared with 1.43), and the best-fitting dust temperature is the same for 89 per cent of the galaxies. The median ratios of the dust luminosity, the FUV attenuation, the SFR averaged over 100\,Myr and the stellar mass between the two fits are 1.000, and the Spearman coefficients between them are 1.00, 0.99, 0.99 and 0.99, although the dust luminosity differs by more than 10 per cent for 15 per cent of the galaxies. The changes are spread evenly between the 84 detections (12 per cent change temperature) and the 714 upper limits (11 per cent), and the reduced $\chi^2$ of the detections falls from 1.97 to 1.65 when the radio band is not fitted, which shows that the fixed $q_{\rm IR}$ is in some tension with their flux densities. The conclusions of this paper therefore do not depend on the radio band once the faint values are treated as upper limits. The fit without radio also gives a prediction of the 1.4\,GHz flux density, from the fitted $L_{\rm dust}$ and $q_{\rm IR}=2.58$, which can be compared with the FIRST data. For the 84 detections the predicted flux density is a median of 0.71 times the observed one (a mean offset of $-0.22$\,dex with a scatter of 0.36\,dex), it is lower than the observed one for 69 per cent of them, and it is within a factor of two for 58 per cent, equivalent to an observed $q_{\rm IR}$ about 0.15\,dex below 2.58. For the 714 upper limits the predicted flux density has a median of 0.52\,mJy and is below the 0.75\,mJy limit for 78 per cent; for the other 156 galaxies (22 per cent) it exceeds the limit, and for 35 it exceeds 1.5\,mJy. Detected galaxies are the radio-brightest for their infrared luminosity, and those which are radio-quiet relative to $q_{\rm IR}=2.58$ are only upper limits, so the two groups are consistent with an intrinsic scatter of a few tenths of a dex in $q_{\rm IR}$ around the assumed value, but with this data we cannot test whether the mean value is correct. The 156 upper-limit galaxies for which the prediction exceeds the limit are nearby (median $z=0.07$, compared with 0.15 for the other upper limits) and have lower dust luminosities (median $\log L_{\rm dust}=37.2$, compared with 37.6, with $L_{\rm dust}$ in W). The fraction of the upper limits for which the prediction exceeds the limit falls from 59 per cent at $z<0.05$ to 31, 15 and 8 per cent for $0.05<z<0.1$, $0.1<z<0.2$ and $0.2<z<0.3$, and to 4 per cent at $z>0.3$, as expected for a fixed flux-density limit, which is a deeper limit in luminosity for nearby galaxies. The prediction implies a lower limit on $q_{\rm IR}$ of $2.58+\log(S_{\rm pred}/0.75\,\mathrm{mJy})$ for each of them (Fig.~\ref{fig:qir_redshift}): the median is 2.70 (16--84 per cent range 2.61--2.91), 29 galaxies have limits above 2.9, and none above 3.25. These galaxies therefore need to be radio-quiet relative to the assumed relation by only 0.1--0.3\,dex in most cases, which is the size of the scatter of the detections. The same selection explains why the observed $q_{\rm IR}$ of the detections decreases with redshift (Fig.~\ref{fig:qir_redshift}): at higher redshifts only galaxies that are bright in the radio for their infrared luminosity exceed the detection limit, so the trend is not evidence of an evolution of $q_{\rm IR}$.
The dust luminosity of \textsc{cigale}, which we use for $L_{\rm IR}$ throughout, is the total luminosity absorbed and re-emitted by the dust, integrated over all wavelengths, whereas the \citet{kennicutt1998star} calibration and the usual definition of $q_{\rm IR}$ refer to 8--1000\,$\mu$m. For the best-fitting \texttt{casey2012} models of our sample 97 per cent of the dust emission (93--99 per cent for 90 per cent of the galaxies) lies in 8--1000\,$\mu$m, so the difference is at most a few hundredths of a dex and we neglect it.

\subsection{Comparison of the SFR indicators}
\label{sec:comparison}

\begin{figure*}
	\includegraphics[width=\textwidth]{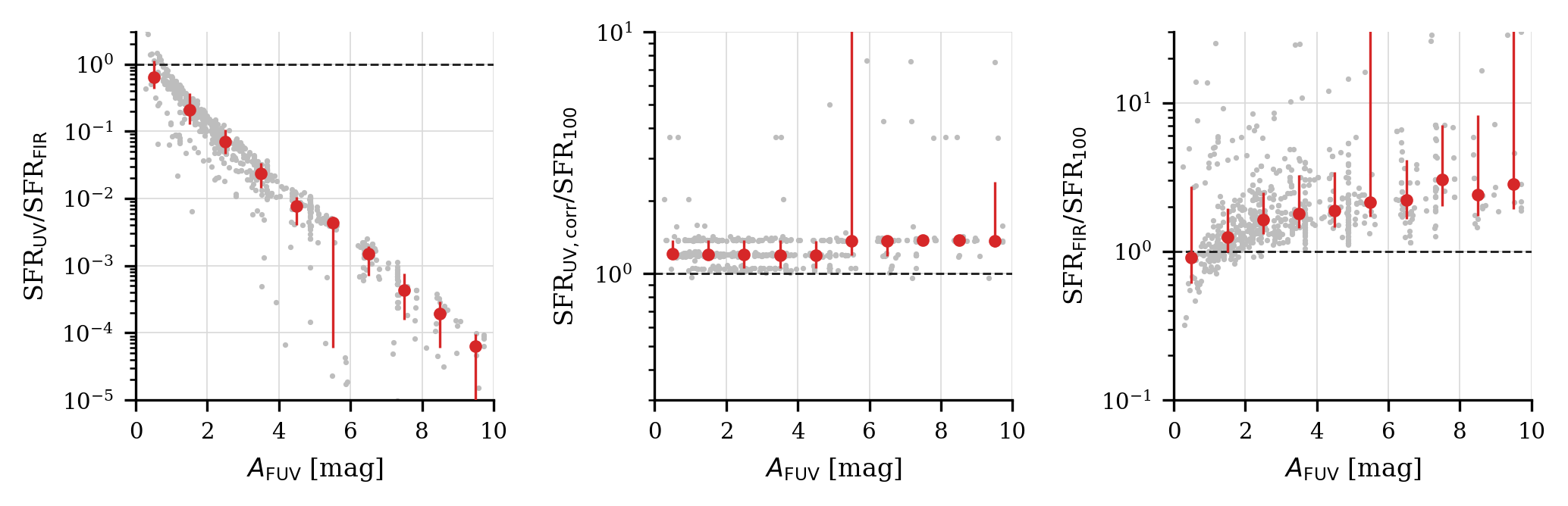}
	\caption{Ratios of single-band SFRs to the FIR SFR (left) or to the SFR of the best-fitting model averaged over 100\,Myr (middle and right), as a function of $A_{\rm FUV}$. Grey points are individual galaxies; the coloured points and bars show the median and the 16--84 per cent range in bins of 1\,mag (bins with fewer than five galaxies are omitted). The dashed line is a ratio of one. In the three panels 12, 12 and 25 galaxies lie outside the vertical range of the plot (in the right panel mostly the galaxies with passive models; Section~\ref{sec:timescales}), and the five galaxies with $A_{\rm FUV}>10$\,mag lie beyond the right-hand edge.}
	\label{fig:sfr_ratios}
	\par\smallskip{\small Alt text: Three panels plotted against FUV attenuation from 0 to 10\,mag, with grey points for individual galaxies and red median points with bars. Left: the ratio of the uncorrected UV SFR to the FIR SFR falls from about 1 to $10^{-4}$ as the attenuation increases. Middle: the attenuation-corrected UV SFR is about 1.2 times the 100-Myr SFR at all attenuations. Right: the ratio of the FIR SFR to the 100-Myr SFR rises from about 1 to 3.}
\end{figure*}

We compare the single-band SFRs of Eqs.~(\ref{eq:sfr_uv}) and~(\ref{eq:sfr_fir}) with the SFR of the best-fitting \textsc{cigale} model averaged over 100\,Myr, $\sfr{100}$, which we take as the reference since it is constrained by the whole SED (the instantaneous and 10\,Myr SFRs differ from it by less than 2 per cent in the median; Section~\ref{sec:timescales}). The radio SFR is not included, for the reason given in Section~\ref{sec:indicators}. Table~\ref{tab:sfr_ratios} and Fig.~\ref{fig:sfr_ratios} show the ratios as a function of $A_{\rm FUV}$.

Without correction for attenuation, the UV SFR is a small fraction of the SFR of the model: the median $\sfr{UV}/\sfr{100}$ is 0.094. The median $\sfr{UV}/\sfr{FIR}$ is 0.25 for $A_{\rm FUV}<2$\,mag, 0.05 for 2--4\,mag, 0.007 for 4--6\,mag and $4\times10^{-4}$ for $A_{\rm FUV}>6$\,mag. Over the whole sample the uncorrected UV SFR is less than half of the FIR SFR for 94 per cent of the galaxies and less than a tenth for 67 per cent; the latter fraction is 9 per cent for $A_{\rm FUV}<2$\,mag, 87 per cent for 2--4\,mag and 100 per cent for $A_{\rm FUV}>4$\,mag. This is the quantitative form of the statement that UV SFRs without an attenuation correction underestimate the obscured star formation in dusty galaxies. When corrected with the fitted $A_{\rm FUV}$, the UV SFR agrees with $\sfr{100}$ to within $\approx40$ per cent in every bin of attenuation (median ratio 1.2--1.4). This is a consistency check on the fit, since the attenuation and the SFR come from the same model, and not an independent test.

The FIR SFR is higher than $\sfr{100}$ by a factor that increases with attenuation: the median ratio is 1.2 for $A_{\rm FUV}<2$\,mag, 1.7 for 2--4\,mag, 2.0 for 4--6\,mag and 2.7 for $A_{\rm FUV}>6$\,mag, and 1.7 overall. Equivalently, the median $\sfr{100}/L_{\rm dust}$ of the models is $2.6\times10^{-37}\,\Msun\,\mathrm{yr^{-1}\,W^{-1}}$, 0.58 of the \citet{kennicutt1998star} value. The trend does not change when the 19 galaxies with passive models (Section~\ref{sec:timescales}), which lie above the plotted range in Fig.~\ref{fig:sfr_ratios}, are excluded (median ratios 1.2, 1.7, 1.9 and 2.5), and we do not know its cause. A possible explanation is that the FIR calibration assumes that the dust is heated by young stars, whereas in the models part of the dust luminosity is powered by older stars, which are attenuated in proportion to the attenuation of the young stars. The best-fitting models support this. At fixed $A_{\rm FUV}$ the ratio $\sfr{FIR}/\sfr{100}$ decreases strongly with the specific SFR of the model (partial Spearman coefficient $-0.84$), as expected if the old stars contribute more of the dust heating in galaxies with less recent star formation, and at fixed specific SFR it still increases with $A_{\rm FUV}$ (partial coefficient 0.83), as expected if a larger fraction of the light of the older stars is absorbed in the more attenuated galaxies. The specific SFR does not decrease with attenuation (median $\log(\sfr{100}/M_*)=-10.2$ for $A_{\rm FUV}<2$\,mag and $-10.0$ for $A_{\rm FUV}>6$\,mag), so the trend with attenuation is not a trend with the star-formation history in disguise. The relative attenuation of the old stars, \texttt{Av\_old\_factor}, increases with $A_{\rm FUV}$ but adds nothing at fixed $A_{\rm FUV}$ (partial coefficient $-0.04$), and neither does the age of the model ($-0.07$). As a check on the fitted attenuations that does not depend on the attenuation law of the fits, we compared them with the attenuation from the infrared excess, $A_{\rm FUV}=2.5\log(1+0.46\,L_{\rm dust}/L_{\rm FUV})$ \citep{hao2011dust}, using the observed FUV flux densities (without $k$-correction) of the 603 galaxies with an FUV measurement. The two agree well: the median difference (fitted minus IRX-based) is $-0.05$\,mag, with an interquartile range of $-0.38$ to $+0.15$\,mag and a Spearman coefficient of 0.94, and the median difference is within 0.2\,mag in each of the three bins below 6\,mag. For the 49 galaxies with $A_{\rm FUV}>6$\,mag the fitted attenuation is higher by a median of 0.9\,mag; their FUV fluxes are faint and uncertain, and these attenuations should be treated with caution.

\begin{table}
	\caption{Ratios of the single-band SFRs to the FIR SFR and to the SFR of the best-fitting model averaged over 100\,Myr, in bins of $A_{\rm FUV}$. $N$ is the number of galaxies, and the other columns are medians, except the fourth, which is the fraction of galaxies. All SFRs are on the Salpeter IMF.}
	\label{tab:sfr_ratios}
	\centering
	\small
	\begin{tabular}{@{}lrcccc@{}}
		\toprule
		$A_{\rm FUV}$ [mag] & $N$ & $\dfrac{\sfr{UV}}{\sfr{FIR}}$ & $\dfrac{\sfr{UV}}{\sfr{FIR}}<0.1$ & $\dfrac{\sfr{UV,corr}}{\sfr{100}}$ & $\dfrac{\sfr{FIR}}{\sfr{100}}$ \\
		\midrule
		0--2 & 247 & 0.25 & 0.09 & 1.20 & 1.16 \\
		2--4 & 296 & 0.047 & 0.87 & 1.20 & 1.73 \\
		4--6 & 129 & 0.0069 & 1.00 & 1.19 & 1.99 \\
		$>6$ & 126 & $5\times10^{-4}$ & 1.00 & 1.37 & 2.66 \\
		\midrule
		All & \Nsamp & --- & 0.67 & 1.20 & 1.72 \\
		\bottomrule
	\end{tabular}
\end{table}

\subsection{SFR timescales}
\label{sec:timescales}

\begin{figure}
	\includegraphics[width=\columnwidth]{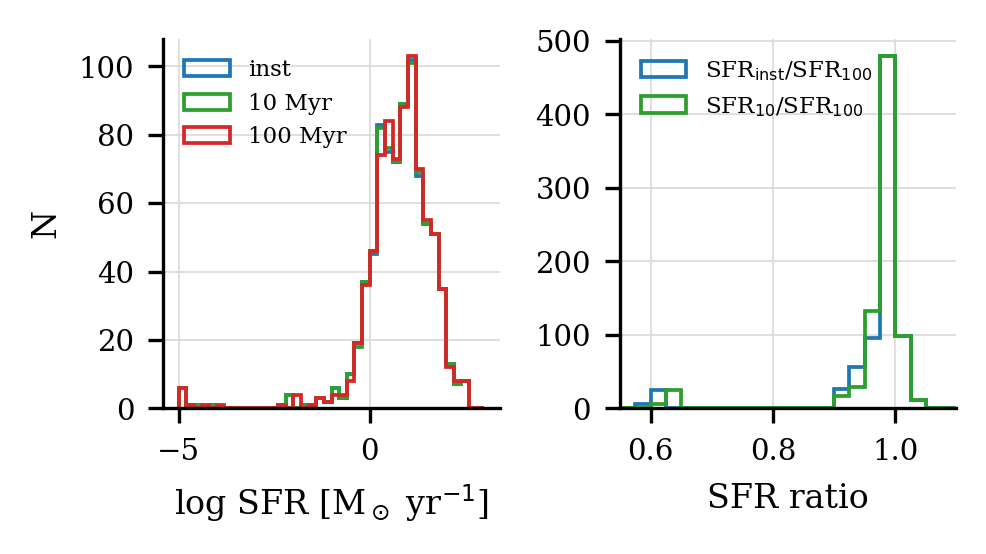}
	\caption{Left: distribution of the logarithm of the SFRs of the best-fitting models on three timescales (instantaneous, 10\,Myr and 100\,Myr; galaxies with SFRs below $10^{-5}\,\Msun\,\mathrm{yr^{-1}}$ are shown in the first bin). Right: distribution of the ratios of the instantaneous and 10\,Myr SFRs to the SFR averaged over 100\,Myr.}
	\label{fig:sfr_timescales}
	\par\smallskip{\small Alt text: Left: almost identical overlapping histograms of the logarithm of the SFR on the three timescales, peaking near 10\,$\Msun\,\mathrm{yr^{-1}}$, with a few galaxies near $10^{-5}\,\Msun\,\mathrm{yr^{-1}}$. Right: histograms of the ratios of the instantaneous and 10-Myr SFRs to the 100-Myr SFR, sharply peaked just below 1, with a small group near 0.6.}
\end{figure}

The SFRs of the best-fitting models on the three timescales are compared in Fig.~\ref{fig:sfr_timescales}. 88 per cent of the galaxies have SFRs below $50\,\Msun\,\mathrm{yr^{-1}}$ on all three timescales, none is above $500\,\Msun\,\mathrm{yr^{-1}}$, and the median SFR is $\approx7.5\,\Msun\,\mathrm{yr^{-1}}$ on all three. The three distributions are almost identical: the median $\sfr{10}/\sfr{100}$ is 0.98 (and the median $\sfr{inst}/\sfr{100}$ is 0.98), and the ratio is within 10 per cent of unity for 96 per cent of the galaxies and within 50 per cent for all of them. This follows from the star formation histories in the grid. A delayed SFH without a burst changes slowly over 100\,Myr unless $\tau$ is short and the age is not much longer, and the best-fitting ages are mostly large (88 per cent are 4\,Gyr or more, 78 per cent are 6\,Gyr or more, and 27 per cent are at the 12\,Gyr edge of the grid). The models therefore cannot produce large differences between SFRs on different timescales, and our comparison cannot show how the tracers respond to recent changes in star formation; this needs models with a recent burst (\texttt{f\_burst}$>0$), which we leave to future work.

A longer averaging period smooths out short-term variations in the star formation, which explains differences between SFRs measured on different timescales \citep{leitherer1999starburst99}. Classical SFR estimators calibrated for a period of 100\,Myr overestimate the SFR by an amount that ranges from around 25 per cent in the FUV to around 65 per cent in the $U$ band, because stars that live for longer than 100\,Myr contribute to the light and the SFR varies on timescales longer than a few tens of Myr; changes in the SFR have little long-term impact but increase the uncertainty in the instantaneous SFR \citep{Boquien2014}.

\begin{figure*}
	\includegraphics[width=\textwidth]{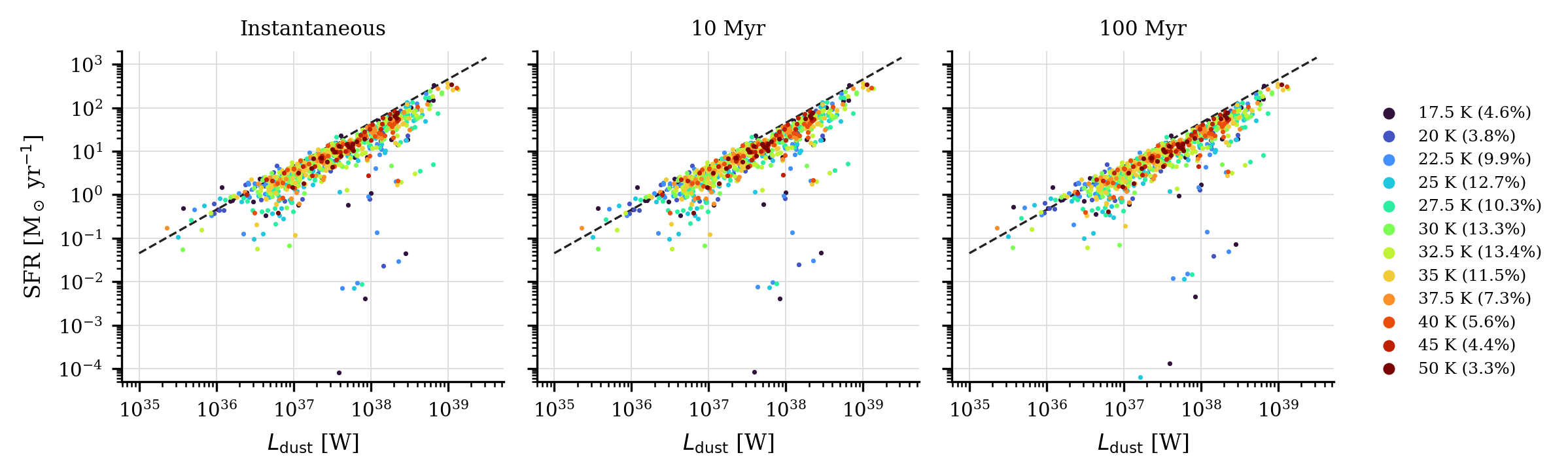}
	\caption{SFR as a function of $L_{\rm dust}$ for all galaxies in the sample, colour-coded by the dust temperature of the best-fitting model (17.5, 20, 22.5, 25, 27.5, 30, 32.5, 35, 37.5, 40, 45 and 50\,K, the twelve values in the grid). The percentage of galaxies at each temperature is given in the legend: 3.3 per cent are at the highest grid value, 50\,K, and 4.6 per cent at the lowest, 17.5\,K. The distribution is similar on all three SFR timescales (left: instantaneous; middle: 10\,Myr; right: 100\,Myr). The dashed line is the \citet{kennicutt1998star} relation, $\mathrm{SFR}=4.5\times10^{-37}\,L_{\rm dust}$ (with $L_{\rm dust}$ in W), which is for a Salpeter IMF like that of the models. Seven or eight galaxies per panel, with SFRs below $5\times10^{-5}\,\Msun\,\mathrm{yr^{-1}}$, lie below the plotted range.}
	\label{fig:Dust Luminosity vs SFR_Subplots.png}
	\par\smallskip{\small Alt text: Three nearly identical logarithmic scatter panels, for the instantaneous, 10-Myr and 100-Myr SFRs, of SFR from $10^{-4}$ to $10^{3}\,\Msun\,\mathrm{yr^{-1}}$ against dust luminosity from $10^{35}$ to $10^{39.5}$\,W, with points coloured by the best-fitting dust temperature from 17.5 to 50\,K. Most galaxies lie in a band just below the dashed \citet{kennicutt1998star} line, and a small group lies far below it.}
\end{figure*}

Fig.~\ref{fig:Dust Luminosity vs SFR_Subplots.png} shows the SFR as a function of the dust luminosity for the three timescales, colour-coded by the dust temperature of the best-fitting model. The twelve colours are the twelve temperatures in our grid (17.5, 20, 22.5, 25, 27.5, 30, 32.5, 35, 37.5, 40, 45 and 50\,K), and the legend gives the percentage of galaxies at each. The SFR increases with $L_{\rm dust}$ over almost four orders of magnitude in $L_{\rm dust}$ ($2\times10^{35}$--$1.4\times10^{39}$\,W), with a slope close to unity on logarithmic axes, and the relation is very similar on the three timescales, as is expected for the smooth SFHs in our grid. At $L_{\rm dust}=10^{38}$\,W the \citet{kennicutt1998star} calibration gives $\approx45\,\Msun\,\mathrm{yr^{-1}}$, which lies within the range occupied by most of the galaxies. The median $\sfr{100}/L_{\rm dust}$ of the best-fitting models is $2.6\times10^{-37}\,\Msun\,\mathrm{yr^{-1}\,W^{-1}}$, 0.58 of the \citet{kennicutt1998star} value. Three features of the figure need comment.

First, the main locus is not a single band. It is divided into several narrow, nearly parallel sequences of unit slope. These are set by the SFH in the model: the combination of \texttt{tau\_main} and \texttt{age\_main} (56 different combinations occur among the best-fitting models) explains 99.6 per cent of the variance of $\log(\sfr{100}/L_{\rm dust})$ between galaxies, whereas the dust temperature explains 4 per cent, and the median ratio in the main locus is the same to within 0.15\,dex for all twelve temperatures. Each sequence corresponds to one model of the grid (a fixed combination of SFH, attenuation and dust parameters, and hence a fixed ratio $\mathrm{SFR}/L_{\rm dust}$) scaled by the stellar mass, so that the spacing of the sequences reflects the sampling of the grid and not the galaxy population, and the relation in this figure is in part determined by the models. The dust temperature carries little information in this figure, and the colours should not be over-interpreted.

Second, the best-fitting dust temperatures are spread over the grid: 17.5\,K: 4.6; 20\,K: 3.8; 22.5\,K: 9.9; 25\,K: 12.7; 27.5\,K: 10.3; 30\,K: 13.3; 32.5\,K: 13.4; 35\,K: 11.5; 37.5\,K: 7.3; 40\,K: 5.6; 45\,K: 4.4; 50\,K: 3.3 per cent, with a median of 30\,K. With the grid of the Thesis (21--40\,K) 81 per cent of the galaxies were best fitted by the lowest temperature, so that pile-up was due to the grid, and, as we show in Section~\ref{sec:radio}, to the treatment of the radio flux densities. The distribution now peaks at 25--35\,K (61 per cent of the galaxies) and has no pile-up at either end of the grid (4.6 per cent at 17.5\,K and 3.3 per cent at 50\,K; with the intermediate grid, 10 per cent were at its upper edge of 40\,K), which we discuss in Section~\ref{sec:discussion}. The temperature is the observed-frame value, because the fits were made at $z=0$ (Section~\ref{sec:redshift}); the rest-frame value is higher by a factor of $\approx(1+z)$, as the fits at the true redshifts confirm (Section~\ref{sec:redshift}), so for the median redshift 30\,K corresponds to $\approx34$\,K. The likelihood-weighted (Bayesian) temperatures from \texttt{pdf\_analysis}, obtained by refitting the sample with the same grid and the dust parameters added to the list of estimated properties (the best-fitting models are identical to those of the original fit), have the same distribution: a median of 30.7\,K, with 68 per cent of the galaxies between 24.7 and 36.4\,K, a median uncertainty of 3.7\,K, and a Spearman correlation of 0.91 with the best-fitting values. For 83 per cent of the galaxies the Bayesian value is within its uncertainty of the best-fitting one (median difference $+0.3$\,K). Because the Bayesian estimate averages over all the models that fit about equally well, it is not restricted to the grid values: the 37 galaxies best fitted at 17.5\,K have a median Bayesian temperature of 19.4\,K (median uncertainty 2.3\,K), so the lowest grid value is not a hard limit for them, whereas the 26 best fitted at 50\,K have 48.9\,K (2.1\,K), so their temperatures are limited by the upper edge of the grid. With \textsc{cigale} v2025.1 the Bayesian temperatures agree with these to within 1\,K for 74 per cent of the galaxies (median difference 0.3\,K, Spearman correlation 0.98). For comparison, isothermal fits of 13\,826 250-$\mu$m-selected H-ATLAS galaxies at $z<0.5$ give an effective temperature of 22--24\,K with $\beta=1.8$--2.0 \citep{smith2013isothermal}, \citet{amblard2010herschel} found an average of $28\pm8$\,K for H-ATLAS sources with optical counterparts detected in at least three FIR bands, and IR-luminous galaxies ($L_{\rm IR}>10^{10}\,L_\odot$) at $0.1<z<2$ have average temperatures of 25--45\,K \citep{symeonidis2013herschel}. Our rest-frame median of $\approx34$\,K is within the range of the IR-luminous galaxies, but warmer than the typical H-ATLAS values. The comparison is only approximate, because the fitted temperature depends on the emissivity index, with which it is degenerate \citep{smith2013isothermal} and which takes all five values of our grid (median 1.8), and on the model, since in \texttt{casey2012} a mid-infrared power law is fitted together with the modified blackbody; our sample is also restricted to the brightest 250-$\mu$m sources of G09 (Section~\ref{sec:selection}).

Third, a small group of 19 galaxies (2.4 per cent) lies more than 1.5\,dex below the median $\sfr{100}/L_{\rm dust}$, on the lowest sequences in the figure. All of them have short \texttt{tau\_main} (100--1000\,Myr) and ages of 1.5--12\,Gyr, so that their current SFR is orders of magnitude below their past average (median specific SFR $10^{-14.2}\,\mathrm{yr^{-1}}$, compared with $10^{-10.2}\,\mathrm{yr^{-1}}$ for the whole sample). They are not poorly fitted (median reduced $\chi^2$ of 1.14, compared with 1.33 for the sample). In these models the dust is heated by old stars, and $\sfr{FIR}$ would overestimate the SFR by orders of magnitude, which would show that FIR SFRs are insensitive to attenuation but not to heating by evolved stars. Alternatively, the grid, which has no burst component, may be unable to represent a mixture of young and old stars in these galaxies. Six of the 25 such galaxies of the Thesis sample, including the two at $z>1$, were among the excluded sources.

\section{Discussion}
\label{sec:discussion}

\subsection{What the comparison does and does not show}

A central aim of this paper was to compare the UV, FIR and radio tracers under uniform modelling, but a uniform model also couples the tracers to each other, and it is important to distinguish what is independent from what is not. The FUV attenuation and the FIR luminosity are linked by the energy balance, so the relations in Figs.~\ref{fig:afuv_redshift}, \ref{fig:FUV vs FIR} and~\ref{fig:NUV vs FIR} test the self-consistency of the fits, and agreement with the empirical IRX--attenuation relations, but not the physical origin of the FIR emission. In the models the radio emission is predicted from $L_{\rm IR}$ with a fixed $q_{\rm IR}$, so a model radio SFR is the FIR SFR by another name. Independent information comes only from the observed data: the observed radio flux density (through the observed $q_{\rm IR}$, or the residual of the fit in the VLA L-band), the observed FUV and NUV fluxes, and the quality of the fits, and from comparison with external calibrations such as \citet{hao2011dust}. Papers that compare SFR tracers need to say clearly, as we have tried to do here, which of their relations are a test and which are a consequence of the model.

\subsection{Timescales}

The tracers sample different parts of the SFH. UV continuum emission is dominated by massive stars with significant lifetimes, so it responds to star formation over tens of Myr; the radio and FIR emission respond to dust heating and supernova remnants, and average star formation over longer timescales, of order 100\,Myr \citep{kennicutt1998star}. This is why UV-based SFRs are reliable, even after correction for attenuation, only for recent star formation, and why FIR and radio SFRs are better for activity averaged over $\sim 100$\,Myr. The star formation of galaxies also varies with time: bursts such as those inferred for `extreme emission-line galaxies', with durations of $\sim 15$\,Myr, are evidence that the SFR variability is common at high redshift \citep{flores2021time}. Galaxies on the main sequence also change their properties over time. \citet{tacchella2016confinement} found that galaxies with low SFR have lower central gas densities and gas fractions and longer depletion times, whereas galaxies with high SFR are compact, with high gas fractions and short depletion times (`blue nuggets'). Variability in the SFR results from cosmological inflows, mergers and recycling outflows on longer and intermediate timescales, and from the life cycle of molecular clouds in the ISM on shorter timescales \citep{10.1093/mnras/staa3893}. A comparison of tracers on a given galaxy therefore reflects its recent SFH as well as any differences in the calibration. In our models, however, this cannot be tested, because the SFHs are smooth and the SFRs on the three timescales differ by less than 2 per cent in the median (Section~\ref{sec:timescales}); the statements about timescales in this section are expectations from the literature and not results of this paper.

\subsection{Limitations of the modelling}

\textit{Redshift.} All galaxies were fitted at $z=0$, with no $k$-corrections, and with distance scaling applied afterwards (Section~\ref{sec:redshift}). The distance scaling restores the luminosities, but the lack of $k$-corrections affects different bands differently (the GALEX and SPIRE bands most), and so it can affect comparisons between the indicators. The fitted dust temperature is an observed-frame value, which is relevant to Fig.~\ref{fig:Dust Luminosity vs SFR_Subplots.png}. The main results do not depend on the high-redshift tail: for the 674 galaxies with $z<0.3$ (84 per cent) the Spearman coefficient between $A_{\rm FUV}$ and $L_{\rm dust}$ is 0.78 (partial coefficient 0.58 at fixed redshift), that with the infrared excess is 0.95, the uncorrected UV SFR is below a tenth of the FIR SFR for 61 per cent, the median ratios of the corrected UV and the FIR SFR to $\sfr{100}$ are 1.20 and 1.64, and the median dust temperature is 30\,K, all close to the values for the whole sample. A fit of the whole sample at its true redshifts (Section~\ref{sec:redshift}) shows that the $z=0$ approximation changes the dust luminosities and SFRs by only about 0.01\,dex in the median, but that it biases the stellar masses high by a median of 0.14\,dex, lowers the dust temperatures by a factor of about $(1+z)$, and overestimates $A_{\rm FUV}$ by a median of 1.1\,mag at $z>0.2$. In that fit 19 per cent of the galaxies are at the upper edge of the temperature grid, and a re-fit at the true redshifts with a temperature grid extending above 50\,K is the main purpose of a follow-up study.

\textit{Radio constraint.} For 89 per cent of the galaxies (714) the 1.4\,GHz flux density is at the level of the FIRST noise, and we entered it as an upper limit (Section~\ref{sec:data}). When the same values were entered as detections with a 15 per cent uncertainty, the fixed $q_{\rm IR}=2.58$ made the radio band a tight and unjustified constraint on $L_{\rm IR}$: the median reduced $\chi^2$ rose from 1.43 to 2.65 and the fraction of galaxies fitted by dust temperatures of 17.5\,K or lower from 4 to 73 per cent (test with the intermediate grid; Section~\ref{sec:radio}). Upper limits remove this bias, but the radio band still constrains the fits of the 84 galaxies with measured flux densities, whose observed $q_{\rm IR}$ (median 2.51) is slightly below the assumed 2.58, and the observed radio--FIR relation can be tested only for them. A fit without the radio band changes the best-fitting dust temperature for only 11 per cent of the galaxies and leaves the median dust luminosity, attenuation, SFR and stellar mass unchanged (test with the intermediate grid; Section~\ref{sec:radio}), so the radio band does not drive the results, even for the 84 galaxies with measured flux densities, for which it lowers the fit quality (reduced $\chi^2$ of 1.97, against 1.65 without radio, in the same test); a fit with $q_{\rm IR}$ as a free parameter is left for future work. The model radio SFR remains the FIR SFR by another name, and the 714 upper limits give no independent information on the SFR.

\textit{AGN.} We excluded the quasars and the galaxies whose mid-infrared emission is dominated by an AGN (Section~\ref{sec:selection}), and fitted the remaining galaxies without an AGN component (Section~\ref{sec:agn}). Low-luminosity and composite AGN, and radio AGN in hosts with normal WISE colours, are not removed by these cuts. Their UV, infrared and radio emission is attributed to star formation in our fits, which leads to overestimates of the SFRs, especially from the radio. To estimate the size of this effect we repeated the fit of the full sample with an AGN component added to our final grid (\texttt{skirtor2016}, with an AGN fraction of 0 or 0.1 and the other AGN parameters as in the tests of \citealt{evans_etal2026}; \textsc{cigale} v2025.1, $3.3\times10^{7}$ models). Only 35 of the \Nsamp\ galaxies (4 per cent) are better fitted with an AGN fraction of 0.1, and for the sample as a whole the median changes in $L_{\rm dust}$, the stellar mass and the SFR are below 0.01\,dex and that in $A_{\rm FUV}$ is below 0.01\,mag, with Spearman coefficients of at least 0.97 between the two fits. The change of \textsc{cigale} version alone alters the properties by less than 0.01\,dex (and the FUV attenuation by less than 0.01\,mag) in the median, although it lowers the reduced $\chi^2$ by about 2 per cent, because the SDSS filter curves differ between the versions. The same AGN component added to our final grid in v2022.1, the version used for our fits, gives the same result for the 50 test galaxies of \citet{evans_etal2026}: five of them are better fitted with an AGN fraction of 0.1, the other 45 have the same best-fitting model as without the AGN component, and for the five the changes ($-0.12$\,dex in $L_{\rm dust}$ and $-0.09$\,dex in the SFR) are similar to those for the full sample. For the 35 galaxies the AGN takes part of the warm-dust emission, and their dust luminosities and SFRs are lower by a median of 0.14 and 0.10\,dex, their FUV attenuations by 0.47\,mag and their dust temperatures by 5\,K. Of the excluded sources, the fit with an AGN component prefers the non-zero AGN fraction for five of the 11 quasars and both of the galaxies with several FIRST components, but for only four of the 22 WISE-selected AGN, so the SED fit is not a substitute for the WISE selection. The exclusion itself does not change the main results (Section~\ref{sec:selection}).

\textit{Star formation histories and IMF.} The best-fitting age is the oldest in the grid (12\,Gyr) for 27 per cent of the galaxies, and 78 per cent have ages of 6\,Gyr or more, so the grid is still restrictive, and ages of 12\,Gyr exceed the age of the Universe for galaxies at $z\gtrsim0.1$. Extending the grid beyond this would make little difference, because the age of the Universe is only 13.8\,Gyr and because the age and $\tau$ are degenerate: the pile-up at the oldest age is a sign that broad-band photometry cannot constrain the ages. The best-fitting $\tau$ is at the edge of the grid for 15 per cent of the galaxies (11 per cent at the maximum of 15\,000\,Myr, which is an almost constant SFR; 19 per cent were at 8000\,Myr with the intermediate grid). Broad-band photometry constrains the ages of the oldest stars poorly. This may lead to errors in the stellar masses and the old-star contribution to the dust heating, and so to errors in the SFR on 100\,Myr timescales and in the attenuation of the UV-bright galaxies. The choice of the Salpeter IMF affects the absolute SFRs, which have to be compared with calibrations on the same IMF.

\textit{Dust temperature grid.} With the extended grid the best-fitting temperatures are spread between 17.5 and 50\,K (median 30\,K), and the pile-up at the lowest value found with the Thesis grid (81 per cent at 21\,K) has gone. The lowest and highest values, 17.5 and 50\,K, hold 4.6 and 3.3 per cent of the galaxies (Fig.~\ref{fig:Dust Luminosity vs SFR_Subplots.png}), so the temperature is a bound only for a small fraction of the sample at $z=0$ (the Bayesian temperatures, median 48.9\,K for the galaxies at 50\,K and 19.4\,K for those at 17.5\,K, show that the upper edge is a real limit and the lower one is not); at the true redshifts the temperatures are higher by a factor of about $(1+z)$, and a grid extending to 60\,K is needed, with which 10 per cent of the test galaxies are still at its upper edge (Section~\ref{sec:redshift}). The temperatures are observed-frame values, and the grid samples the temperature coarsely, so the SFR--$L_{\rm dust}$ relation is organised into sequences that probably correspond to individual grid models; the temperature-dependent structure in that figure should not be over-interpreted.

\textit{Edges of the parameter grid.} The dust temperature is not the only parameter for which the best-fitting model lies at an edge of the grid (Fig.~\ref{fig:grid_edges}): \texttt{Av\_old\_factor}\,=\,0.1 or 1.0 for 35 per cent of the galaxies (26 per cent at 1.0, the physical maximum, for which the old stars are attenuated as much as the young ones, so that the pile-up cannot be removed by extending the grid), $\beta=1.0$ or 2.5 for 47 per cent (a degeneracy with the temperature, which the FIR photometry cannot break, and not a limit of the grid), \texttt{age\_main}\,=\,1000 or 12\,000\,Myr for 30 per cent (27 per cent at the maximum), \texttt{tau\_main}\,=\,100 or 15\,000\,Myr for 15 per cent, $\alpha=1.5$ or 3.0 for 6 per cent, \texttt{Av\_young}\,=\,0.25 or 8\,mag for 3 per cent and $T=17.5$ or 50\,K for 8 per cent. Extending the grid (Section~\ref{sec:cigale}) reduced the fractions at the upper edge for $T$ (from 10 to 3.3 per cent), $\alpha$ (from 13 to 2 per cent) and $\tau$ (from 19 to 11 per cent), but not for the age or \texttt{Av\_old\_factor}. The median dust luminosity and SFR are unchanged by the extension (the median ratio of the two fits is 1.00 for both), so the main results do not depend on it, although the values for individual galaxies changed by up to a factor of several. At least one of the six parameters in Fig.~\ref{fig:grid_edges} is at an edge for 79 per cent of the galaxies. For these galaxies the corresponding quantity is a bound and not a measurement, and the attenuations, ages and temperatures derived from them should be treated as such.

\textit{Dust emission model.} \texttt{casey2012} gives the dust temperature but not PAH emission, and fits our galaxies less well than the Draine \& Li models (Section~\ref{sec:cigale}); the dust luminosity and attenuation could be affected by this choice. We therefore repeated the fit of the full sample with the best \texttt{dl2007} grid of \citet{evans_etal2026}, keeping everything else, including the radio module and upper limits, unchanged (Section~\ref{sec:cigale}). For the \Nsamp\ galaxies the two fits give almost the same values: $L_{\rm dust}$ is higher with \texttt{dl2007} by a median of 0.01\,dex (interquartile range $-0.04$ to $+0.07$\,dex; Spearman rank coefficient 0.90), $A_{\rm FUV}$ differs by a median of 0.00\,mag (interquartile range $-0.14$ to $+0.20$\,mag, for a median $A_{\rm FUV}$ of 2.8\,mag; Spearman 0.98), the stellar mass by a median of 0.01\,dex (interquartile range $-0.12$ to $+0.07$\,dex) and the SFR by 0.04\,dex (interquartile range 0.00 to $+0.08$\,dex; Spearman 0.96). On the 50 test galaxies a \texttt{dl2014} grid gives a similar result ($L_{\rm dust}$ 0.01\,dex lower than with \texttt{casey2012}, and the same median $A_{\rm FUV}$). The choice of dust-emission model therefore matters for the fit quality and the dust temperature, but little for the quantities used to correct the UV SFRs.

\textit{Selection.} Our sample is drawn from the brightest 250-$\mu$m sources in the G09 field (Section~\ref{sec:selection}), and the SPIRE selection of H-ATLAS favours dusty, FIR-bright galaxies; the requirement of multiwavelength data may add further biases. Our results should therefore not be extended to fainter FIR-selected samples or to UV-selected samples, where the fraction of galaxies with low attenuation will be higher.

\subsection{Running \textsc{cigale} on a modest computer}

The fits in this paper show what is possible without HPC access, and also what is lost. A complete UV-to-radio energy-balance fit of the 852 galaxies of the Thesis sample with $1.63\times10^{7}$ models per galaxy took 1\,h\,38\,min on six cores of a mini PC, and a grid of this size is within the memory limit of a laptop with 8\,GB of RAM. Three choices made this possible: fixing the redshift at zero, which reduces the number of models by a factor of 64 for our sample and puts the burden of the correction on a distance scaling applied afterwards; omitting the AGN component, which we could do only by excluding the galaxies with a dominant AGN from the sample; and limiting the number of values of each parameter, so that the total stayed below the $\sim1.5$--$1.8\times10^{7}$ models above which \textsc{cigale} ran out of memory on the laptop. Each of these has a cost, which we have tried to quantify: the lack of $k$-corrections (Section~\ref{sec:redshift}), the smaller sample and the attribution of any remaining AGN emission to star formation (Section~\ref{sec:agn}), and the pile-up of best-fitting parameters at the edges of the grid (Fig.~\ref{fig:grid_edges}). In practice a user in this situation should check what fraction of the best-fitting values lie at the edges of each grid, change the values of the parameters for which this fraction is large (dropping values that are never chosen to make room for new ones), and prefer quantities such as the dust luminosity and the attenuation relative to the infrared excess, which are better constrained by the data, to those that are bounded by the grid.

Fixing the redshift is the choice with the largest effect. When the fits were made, a fit at the true redshifts with our grid was expected to take about 8 days on the mini PC, from a first test that, as we found later, was probably slowed by memory pressure (Section~\ref{sec:redshift}). With the grid divided into blocks small enough to fit in the memory, the fit of the \Nsamp\ galaxies at their true redshifts took 12.2\,h on the same machine, nine times as long as at $z=0$. A single fit of this kind is therefore practical on a desktop computer with 32\,GB of RAM, but a study that compares many grids, as ours did, needs many such runs, each exposed to power cuts. Fits at the true redshifts with an extended temperature grid and with an AGN component, which increases the number of models further, are the subject of a follow-up study, for which access to HPC resources has been requested.

\subsection{Implications for surveys}

Our results are relevant to surveys that rely on the rest-frame UV, for example at high redshift: where the attenuation is large ($A_{\rm FUV}\gtrsim2$\,mag), the UV flux is reduced by a factor of 6 or more, and an uncorrected SFR can be very wrong. Wherever possible, UV SFRs need to be corrected with the FIR (or a good attenuation estimate), and the FIR and radio can be used as a cross-check of each other, provided that their relationship is tested against observations and not against a model.

\section{Conclusions}
\label{sec:conclusions}

We have compared UV, FIR and radio SFR diagnostics for \Nsamp\ H-ATLAS DR1 galaxies with reliable redshifts, after excluding quasars and galaxies whose mid-infrared emission is dominated by an AGN, using uniform energy-balance SED fitting with \textsc{cigale}. Our main conclusions are as follows.
\begin{enumerate}[label=(\arabic*)]
	\item The FUV attenuation correlates strongly with the FIR luminosity (Spearman $\rho=0.84$), but much of this is due to the dependence of both quantities on redshift in a flux-limited sample (partial $\rho=0.57$ at fixed redshift). The attenuation correlates more tightly with the infrared excess, which does not depend on distance ($\rho=0.96$), as expected for dust-reprocessed star formation; this is partly built into the energy-balance modelling, and demonstrates the self-consistency of the fits.
	\item For $A_{\rm FUV}>2$\,mag the UV flux is reduced by a factor of more than 6, so UV SFRs are unreliable without correction for attenuation: the uncorrected UV SFR is below a tenth of the FIR SFR for 67 per cent of the galaxies, and for all of those with $A_{\rm FUV}>4$\,mag. Corrected with the fitted attenuation it agrees with the model SFR to within $\approx40$ per cent, which is a consistency check on the fit, whereas the FIR SFR overestimates the model SFR by a factor that grows from 1.2 to 2.7 with increasing attenuation (Table~\ref{tab:sfr_ratios}).
	\item For 714 of the \Nsamp\ galaxies the FIRST flux density is at the noise level and must be entered as an upper limit: entered as detections, together with the fixed $q_{\rm IR}=2.58$, they push the fits to low dust temperatures (73 per cent at 17.5\,K or lower, compared with 4 per cent) and a poorer fit (median reduced $\chi^2$ of 2.65, compared with 1.43, in a test with the intermediate grid). For the 84 galaxies with a measured flux density the observed $q_{\rm IR}$ has a median of 2.51 and a scatter of 0.34\,dex, slightly below the assumed value (seven of them have a radio excess of more than 0.5\,dex and are probable radio AGN), and the observed radio and FIR luminosities are correlated (Spearman $\rho=0.89$, 0.48 at fixed redshift). Since the model radio emission is tied to $L_{\rm IR}$, the model radio SFR is the FIR SFR by another name. A fit without the radio band gives the same results for the median dust luminosity, attenuation, SFR and stellar mass, and the dust temperature is unchanged for 89 per cent of the galaxies.
	\item The SFRs of the models on the three timescales differ by less than 2 per cent in the median, because of the smooth SFHs in our grid, so the timescale sensitivity of the tracers cannot be tested with these models.
	\item Fitting all galaxies at $z=0$ changes the dust luminosities and SFRs by only a few hundredths of a dex, but it lowers the dust temperatures by a factor of about $(1+z)$, biases the stellar masses high by a median of 0.14\,dex (0.33\,dex at $z>0.2$), and overestimates the FUV attenuation at $z>0.2$ by a median of 1.1\,mag (from a fit of the whole sample at the true redshifts). Quasars and mid-infrared AGN are excluded; a fit with an AGN component prefers a non-zero AGN fraction for only 4 per cent of the remaining galaxies, and does not change the results for the sample as a whole. With the extended grid the best-fitting dust temperatures are spread between 17.5 and 50\,K (median 30\,K), with no pile-up at either end, but the best-fitting values still lie at an edge of the grid for several other parameters (\texttt{Av\_old\_factor} for 35 per cent of the galaxies, $\beta$ for 47 per cent, the age for 30 per cent, $\tau$ for 15 per cent), most of which are physical limits or degeneracies.
	\item The fits use \textsc{cigale} v2022.1 and $1.63\times10^{7}$ models per galaxy, fixed redshifts and no AGN component. The complete fit of the 852 galaxies of the Thesis sample took 1\,h\,38\,min on a mini PC, which was enough to compare the UV, FIR and radio tracers without HPC access; the limit on the number of models restricts the resolution of the parameter grid, and is the main reason for the limitations above.
\end{enumerate}
In a follow-up study we will re-fit the sample at the true redshifts with a temperature grid extending to at least 60\,K and an AGN component, and fit with and without the radio band, so that the radio flux density can be used as an independent test of the FIR--radio relation and to calibrate the three diagnostics against each other.

\section*{Acknowledgements}

We thank the BIUST Astrophysics group for their support, and BIUST for the laptop on which the Thesis fits were made. This work is based on observations from \textit{Herschel}, GALEX, SDSS and FIRST. GAMA is a joint European-Australasian project based around a spectroscopic campaign using the Anglo-Australian Telescope. The GAMA input catalogue is based on data taken from the Sloan Digital Sky Survey and the UKIRT Infrared Deep Sky Survey. Complementary imaging of the GAMA regions is being obtained by a number of independent survey programmes including GALEX MIS, VST KiDS, VISTA VIKING, WISE, \textit{Herschel}-ATLAS, GMRT and ASKAP providing UV to radio coverage. GAMA is funded by the STFC (UK), the ARC (Australia), the AAO, and the participating institutions. The GAMA website is \url{http://www.gama-survey.org/}. This publication has made use of data from the VIKING survey from VISTA at the ESO Paranal Observatory, programme ID 179.A-2004. Data processing has been contributed by the VISTA Data Flow System at CASU, Cambridge and WFAU, Edinburgh. This publication makes use of data products from the Wide-field Infrared Survey Explorer, which is a joint project of the University of California, Los Angeles, and the Jet Propulsion Laboratory/California Institute of Technology, and NEOWISE, which is a project of the Jet Propulsion Laboratory/California Institute of Technology. WISE and NEOWISE are funded by the National Aeronautics and Space Administration. The FIRST and NVSS surveys were made with the Very Large Array of the National Radio Astronomy Observatory. The National Radio Astronomy Observatory and Green Bank Observatory are facilities of the U.S. National Science Foundation operated under cooperative agreement by Associated Universities, Inc.

We used Claude, a large language model developed by Anthropic, as an assistant in preparing this paper: to help write and run the \textsc{python} scripts that compute the statistics and make the figures from the \textsc{cigale} outputs, to check the numbers quoted in the text against the fit results, and to edit the text and draft the descriptions of the figures. The authors directed this work, checked its results, and take full responsibility for the content of the paper.

\section*{Data Availability}

The data used in this paper are from the public H-ATLAS DR1 release (\url{https://www.h-atlas.org/public-data/download}; \citealt{valiante2016herschel, bourne2016herschel}) and the public GALEX, SDSS, NVSS and FIRST archives. The \textsc{cigale} configuration is given in Appendix~\ref{app:cigale}. The derived catalogue of fitted properties will be shared on reasonable request to the corresponding author.

\bibliographystyle{mnras}
\bibliography{References_v1, refs_new_v1}

@article{dutkowska2023star,
	author = {Dutkowska, KM and Kristensen, LE},
	journal = {Astronomy \& Astrophysics},
	pages = {A95},
	publisher = {EDP Sciences},
	title = {Star-formation-rate estimates from water emission},
	volume = {674},
	year = {2023}}

@article{boquien2019cigale,
	author = {Boquien, M and Burgarella, D and Roehlly, Y and Buat, V and Ciesla, L and Corre, D and Inoue, AK and Salas, H},
	journal = {Astronomy \& Astrophysics},
	pages = {A103},
	publisher = {EDP Sciences},
	title = {CIGALE: a Python code investigating galaxy emission},
	volume = {622},
	year = {2019}}

@article{lozano2023model,
	author = {Lozano, Ezequiel and Scannapieco, Cecilia and Nuza, Sebastian E},
	journal = {arXiv preprint arXiv:2307.03341},
	title = {A model for molecular hydrogen-dependent star formation in simulations of galaxy evolution},
	year = {2023}}

@phdthesis{battisti2017ultraviolet,
	author = {Battisti, Andrew},
	school = {University of Massachusetts Libraries},
	title = {Ultraviolet to Infrared Star Formation Rate Tracers: Characterizing Dust Attenuation and Emission},
	year = {2017}}

@article{shi2013star,
	author = {Shi, F and Kong, X},
	journal = {arXiv preprint arXiv:1312.1781},
	title = {Star Formation Rate Indicators in WISE/SDSS},
	year = {2013}}

@article{Ot_Floranes_2010,
	author = {Ot{\'\i}-Floranes, H. and Mas-Hesse, J. M.},
	doi = {10.1051/0004-6361/200913384},
	issn = {1432-0746},
	journal = {Astronomy and Astrophysics},
	month = {feb},
	pages = {A61},
	publisher = {EDP Sciences},
	title = {Calibration of star formation rate tracers for short- and long-lived star formation episodes},
	url = {http://dx.doi.org/10.1051/0004-6361/200913384},
	volume = {511},
	year = {2010}}

@article{kennicutt1998star,
	author = {Kennicutt Jr, Robert C},
	journal = {Annual Review of Astronomy and Astrophysics},
	number = {1},
	pages = {189--231},
	publisher = {Annual Reviews 4139 El Camino Way, PO Box 10139, Palo Alto, CA 94303-0139, USA},
	title = {Star formation in galaxies along the Hubble sequence},
	volume = {36},
	year = {1998}}

@article{werhahn2021cosmic,
	author = {Werhahn, Maria and Pfrommer, Christoph and Girichidis, Philipp},
	journal = {Monthly Notices of the Royal Astronomical Society},
	number = {3},
	pages = {4072--4095},
	publisher = {Oxford University Press},
	title = {Cosmic rays and non-thermal emission in simulated galaxies--III. Probing cosmic-ray calorimetry with radio spectra and the FIR--radio correlation},
	volume = {508},
	year = {2021}}

@article{condon1992radio,
	author = {Condon, JJ},
	journal = {In: Annual review of astronomy and astrophysics. Vol. 30 (A93-25826 09-90), p. 575-611.},
	pages = {575--611},
	title = {Radio emission from normal galaxies},
	volume = {30},
	year = {1992}}

@article{magnelli2015far,
	author = {Magnelli, B and Ivison, RJ and Lutz, D and Valtchanov, I and Farrah, D and Berta, Stefano and Bertoldi, F and Bock, J and Cooray, A and Ibar, E and others},
	journal = {Astronomy \& Astrophysics},
	pages = {A45},
	publisher = {EDP Sciences},
	title = {The far-infrared/radio correlation and radio spectral index of galaxies in the SFR--M∗ plane up to z\~{} 2},
	volume = {573},
	year = {2015}}

@article{boquien2016towards,
	author = {Boquien, M and Kennicutt, R and Calzetti, D and Dale, D and Galametz, M and Sauvage, M and Croxall, K and Draine, B and Kirkpatrick, A and Kumari, N and others},
	journal = {Astronomy \& Astrophysics},
	pages = {A6},
	publisher = {EDP Sciences},
	title = {Towards universal hybrid star formation rate estimators},
	volume = {591},
	year = {2016}}

@article{valiante2016herschel,
	author = {Valiante, E and Smith, MWL and Eales, S and Maddox, SJ and Ibar, E and Hopwood, R and Dunne, L and Cigan, PJ and Dye, S and Pascale, Enzo and others},
	journal = {Monthly Notices of the Royal Astronomical Society},
	number = {3},
	pages = {3146--3179},
	publisher = {Oxford University Press},
	title = {The Herschel-ATLAS data release 1--I. Maps, catalogues and number counts},
	volume = {462},
	year = {2016}}

@article{yu2024exploring,
	author = {Yu, Yaosong and Chen, Qihang and Jing, Liang and Pappalardo, Ciro and Miranda, Henrique},
	journal = {arXiv preprint arXiv:2404.07744},
	title = {Exploring the Current Star Formation Rate and Nebula Ratio of Star-Formation Galaxies at z< 0.4 with FADO},
	year = {2024}}

@article{casey2012far,
	author = {Casey, Caitlin M},
	journal = {Monthly Notices of the Royal Astronomical Society},
	number = {4},
	pages = {3094--3103},
	publisher = {Blackwell Science Ltd Oxford, UK},
	title = {Far-infrared spectral energy distribution fitting for galaxies near and far},
	volume = {425},
	year = {2012}}

@article{viaene2016bolometric,
	author = {Viaene, S{\'e}bastien and Baes, Maarten and Bendo, G and Boquien, M and Boselli, A and Ciesla, L and Cortese, Luca and De Looze, Ilse and Eales, S and Fritz, Jacopo and others},
	journal = {Astronomy \& Astrophysics},
	pages = {A13},
	publisher = {EDP Sciences},
	title = {The bolometric and UV attenuation in normal spiral galaxies of the Herschel Reference Survey},
	volume = {586},
	year = {2016}}

@article{hao2011dust,
	author = {Hao, Cai-Na and Kennicutt, Robert C and Johnson, Benjamin D and Calzetti, Daniela and Dale, Daniel A and Moustakas, John},
	journal = {The Astrophysical Journal},
	number = {2},
	pages = {124},
	publisher = {IOP Publishing},
	title = {Dust-corrected star formation rates of galaxies. II. Combinations of ultraviolet and infrared tracers},
	volume = {741},
	year = {2011}}

@article{Casey_2014,
	author = {Casey, C. M. and Scoville, N. Z. and Sanders, D. B. and Lee, N. and Cooray, A. and Finkelstein, S. L. and Capak, P. and Conley, A. and De Zotti, G. and Farrah, D. and Fu, H. and Le Floc'h, E. and Ilbert, O. and Ivison, R. J. and Takeuchi, T. T.},
	doi = {10.1088/0004-637X/796/2/95},
	journal = {The Astrophysical Journal},
	month = {nov},
	number = {2},
	pages = {95},
	publisher = {The American Astronomical Society},
	title = {ARE DUSTY GALAXIES BLUE? INSIGHTS ON UV ATTENUATION FROM DUST-SELECTED GALAXIES},
	url = {https://dx.doi.org/10.1088/0004-637X/796/2/95},
	volume = {796},
	year = {2014}}

@article{10.1111/j.1365-2966.2005.09131.x,
	author = {Burgarella, D. and Buat, V. and Iglesias-P{\'a}ramo, J.},
	doi = {10.1111/j.1365-2966.2005.09131.x},
	eprint = {https://academic.oup.com/mnras/article-pdf/360/4/1413/4885680/360-4-1413.pdf},
	issn = {0035-8711},
	journal = {Monthly Notices of the Royal Astronomical Society},
	month = {07},
	number = {4},
	pages = {1413-1425},
	title = {Star formation and dust attenuation properties in galaxies from a statistical ultraviolet-to-far-infrared analysis},
	url = {https://doi.org/10.1111/j.1365-2966.2005.09131.x},
	volume = {360},
	year = {2005}}

@article{calzetti2000dust,
	author = {Calzetti, Daniela and Armus, Lee and Bohlin, Ralph C and Kinney, Anne L and Koornneef, Jan and Storchi-Bergmann, Thaisa},
	journal = {The Astrophysical Journal},
	number = {2},
	pages = {682},
	publisher = {IOP Publishing},
	title = {The Dust Content and Opacity of ActivelyStar-forming Galaxies},
	volume = {533},
	year = {2000}}

@article{read2018far,
	author = {Read, SHAUN CONISBEE and Smith, DJB and G{\"u}rkan, G and Hardcastle, MJ and Williams, Wendy Louise and Best, PN and Brinks, Elias and Calistro-Rivera, Gabriela and Chy{\.Z}y, KT and Duncan, K and others},
	journal = {Monthly Notices of the Royal Astronomical Society},
	number = {4},
	pages = {5625--5644},
	publisher = {Oxford University Press},
	title = {The Far-Infrared Radio Correlation at low radio frequency with LOFAR/H-ATLAS},
	volume = {480},
	year = {2018}}

@article{lacki2010physics,
	author = {Lacki, Brian C and Thompson, Todd A and Quataert, Eliot},
	journal = {The Astrophysical Journal},
	number = {1},
	pages = {1},
	publisher = {IOP Publishing},
	title = {THE PHYSICS OF THE FAR-INFRARED--RADIO CORRELATION. I. CALORIMETRY, CONSPIRACY, AND IMPLICATIONS},
	volume = {717},
	year = {2010}}

@article{molnar2021non,
	author = {Moln{\'a}r, D{\'a}niel Cs and Sargent, Mark T and Leslie, Sarah and Magnelli, Benjamin and Schinnerer, Eva and Zamorani, Giovanni and Delhaize, Jacinta and Smol{\v{c}}i{\'c}, Vernesa and Tisani{\'c}, Kre{\v{s}}imir and Vardoulaki, Eleni},
	journal = {Monthly Notices of the Royal Astronomical Society},
	number = {1},
	pages = {118--145},
	publisher = {Oxford University Press},
	title = {The non-linear infrared-radio correlation of low-z galaxies: implications for redshift evolution, a new radio SFR recipe, and how to minimize selection bias},
	volume = {504},
	year = {2021}}

@article{Becker1995,
	author = {Becker, R. H. and White, R. L. and Helfand, D. J.},
	doi = {10.1086/176166},
	journal = {The Astrophysical Journal},
	pages = {559},
	title = {The FIRST Survey: Faint Images of the Radio Sky at Twenty Centimeters},
	volume = {450},
	year = {1995}}

@article{Yun_2001,
	author = {Yun, Min S. and Reddy, Naveen A. and Condon, J. J.},
	doi = {10.1086/323145},
	journal = {The Astrophysical Journal},
	month = {jun},
	number = {2},
	pages = {803},
	title = {Radio Properties of Infrared-selected Galaxies in the IRAS 2 Jy Sample},
	url = {https://dx.doi.org/10.1086/323145},
	volume = {554},
	year = {2001}}

@article{cox1988correlation,
	author = {Cox, MJ and Eales, SAE and Alexander, P and Fitt, AJ},
	journal = {Monthly Notices of the Royal Astronomical Society},
	number = {4},
	pages = {1227--1243},
	publisher = {Oxford University Press},
	title = {The correlation between far-infrared and non-thermal radio emission at 151 MHz for galaxies: observations and modelling},
	volume = {235},
	year = {1988}}

@article{flores2021time,
	author = {Flores Vel{\'a}zquez, Jos{\'e} A and Gurvich, Alexander B and Faucher-Gigu{\`e}re, Claude-Andr{\'e} and Bullock, James S and Starkenburg, Tjitske K and Moreno, Jorge and Lazar, Alexandres and Mercado, Francisco J and Stern, Jonathan and Sparre, Martin and others},
	journal = {Monthly Notices of the Royal Astronomical Society},
	number = {4},
	pages = {4812--4824},
	publisher = {Oxford University Press},
	title = {The time-scales probed by star formation rate indicators for realistic, bursty star formation histories from the FIRE simulations},
	volume = {501},
	year = {2021}}

@article{tacchella2016confinement,
	author = {Tacchella, Sandro and Dekel, Avishai and Carollo, C Marcella and Ceverino, Daniel and DeGraf, Colin and Lapiner, Sharon and Mandelker, Nir and Primack Joel, R},
	journal = {Monthly Notices of the Royal Astronomical Society},
	number = {3},
	pages = {2790--2813},
	publisher = {Oxford University Press},
	title = {The confinement of star-forming galaxies into a main sequence through episodes of gas compaction, depletion and replenishment},
	volume = {457},
	year = {2016}}

@article{eales2010herschel,
	author = {Eales, Stephen and Dunne, Loretta and Clements, D and Cooray, A and De Zotti, Gianfranco and Dye, Simon and Ivison, R and Jarvis, M and Lagache, G and Maddox, Steve and others},
	journal = {Publications of the Astronomical Society of the Pacific},
	number = {891},
	pages = {499},
	publisher = {IOP Publishing},
	title = {The herschel atlas},
	volume = {122},
	year = {2010}}

@article{maraston2005evolutionary,
	author = {Maraston, Claudia},
	journal = {Monthly Notices of the Royal Astronomical Society},
	number = {3},
	pages = {799--825},
	publisher = {Blackwell Science Ltd Oxford, UK},
	title = {Evolutionary population synthesis: models, analysis of the ingredients and application to high-z galaxies},
	volume = {362},
	year = {2005}}

@article{bruzual2003stellar,
	author = {Bruzual, Gustavo and Charlot, Stephane},
	journal = {Monthly Notices of the Royal Astronomical Society},
	number = {4},
	pages = {1000--1028},
	publisher = {Blackwell Science Ltd Oxford, UK},
	title = {Stellar population synthesis at the resolution of 2003},
	volume = {344},
	year = {2003}}

@article{Markov2024,
	author = {Markov, V. and Gallerani, S. and Ferrara, A. and Pallottini, A. and Parlanti, E. and Di Mascia, F. and Sommovigo, L. and Kohandel, M.},
	journal = {arXiv preprint arXiv:2402.05996},
	title = {Dust attenuation evolution in $ z \sim 2$-$12 $ JWST galaxies},
	url = {https://arxiv.org/abs/2402.05996},
	year = {2024}}

@article{planck2020,
	author = {{Planck Collaboration} and Aghanim, N. and Akrami, Y. and Ashdown, M. and Aumont, J. and Baccigalupi, C. and Ballardini, M. and Banday, A. J. and Barreiro, R. B. and Bartolo, N. and Basak, S. and Battye, R. and Benabed, K. and Bernard, J.-P. and Bersanelli, M. and Bielewicz, P. and Bock, J. J. and Bond, J. R. and Borrill, J. and Bouchet, F. R. and Boulanger, F. and Bucher, M. and Burigana, C. and Butler, R. C. and Calabrese, E. and Cardoso, J.-F. and Carron, J. and Challinor, A. and Chiang, H. C. and Chluba, J. and Colombo, L. P. L. and Combet, C. and Contreras, D. and Crill, B. P. and Cuttaia, F. and de Bernardis, P. and de Zotti, G. and Delabrouille, J. and Delouis, J.-M. and Di Valentino, E. and Diego, J. M. and Dor{\'e}, O. and Douspis, M. and Ducout, A. and Dupac, X. and Dusini, S. and Efstathiou, G. and Elsner, F. and En{\ss}lin, T. A. and Eriksen, H. K. and Fantaye, Y. and Farhang, M. and Fergusson, J. and Fernandez-Cobos, R. and Finelli, F. and Forastieri, F. and Frailis, M. and Fraisse, A. A. and Franceschi, E. and Frolov, A. and Galeotta, S. and Galli, S. and Ganga, K. and G{\'e}nova-Santos, R. T. and Gerbino, M. and Ghosh, T. and Gonz{\'a}lez-Nuevo, J. and G{\'o}rski, K. M. and Gratton, S. and Gruppuso, A. and Gudmundsson, J. E. and Hamann, J. and Handley, W. and Hansen, F. K. and Herranz, D. and Hildebrandt, S. R. and Hivon, E. and Huang, Z. and Jaffe, A. H. and Jones, W. C. and Karakci, A. and Keih{\"a}nen, E. and Keskitalo, R. and Kiiveri, K. and Kim, J. and Kisner, T. S. and Knox, L. and Krachmalnicoff, N. and Kunz, M. and Kurki-Suonio, H. and Lagache, G. and Lamarre, J.-M. and Lasenby, A. and Lattanzi, M. and Lawrence, C. R. and Le Jeune, M. and Lemos, P. and Lesgourgues, J. and Levrier, F. and Lewis, A. and Liguori, M. and Lilje, P. B. and Lilley, M. and Lindholm, V. and L{\'o}pez-Caniego, M. and Lubin, P. M. and Ma, Y.-Z. and Mac{\'\i}as-P{\'e}rez, J. F. and Maggio, G. and Maino, D. and Mandolesi, N. and Mangilli, A. and Marcos-Caballero, A. and Maris, M. and Martin, P. G. and Martinelli, M. and Mart{\'\i}nez-Gonz{\'a}lez, E. and Matarrese, S. and Mauri, N. and McEwen, J. D. and Meinhold, P. R. and Melchiorri, A. and Mennella, A. and Migliaccio, M. and Millea, M. and Mitra, S. and Miville-Desch{\^e}nes, M.-A. and Molinari, D. and Montier, L. and Morgante, G. and Moss, A. and Natoli, P. and N{\o}rgaard-Nielsen, H. U. and Pagano, L. and Paoletti, D. and Partridge, B. and Patanchon, G. and Peiris, H. V. and Perrotta, F. and Pettorino, V. and Piacentini, F. and Polastri, L. and Polenta, G. and Puget, J.-L. and Rachen, J. P. and Reinecke, M. and Remazeilles, M. and Renzi, A. and Rocha, G. and Rosset, C. and Roudier, G. and Rubi{\~n}o-Mart{\'\i}n, J. A. and Ruiz-Granados, B. and Salvati, L. and Sandri, M. and Savelainen, M. and Scott, D. and Shellard, E. P. S. and Sirignano, C. and Sirri, G. and Spencer, L. D. and Sunyaev, R. and Suur-Uski, A.-S. and Tauber, J. A. and Tavagnacco, D. and Tenti, M. and Toffolatti, L. and Tomasi, M. and Trombetti, T. and Valenziano, L. and Valiviita, J. and Van Tent, B. and Vibert, L. and Vielva, P. and Villa, F. and Vittorio, N. and Wandelt, B. D. and Wehus, I. K. and White, M. and White, S. D. M. and Zacchei, A. and Zonca, A.},
	journal = {Astronomy \& Astrophysics},
	number = {A6},
	title = {Planck 2018 results. VI. Cosmological parameters},
	volume = {641},
	year = {2020}}

@article{inoue2011rest,
	author = {Inoue, Akio K},
	journal = {Monthly Notices of the Royal Astronomical Society},
	number = {3},
	pages = {2920--2931},
	publisher = {Blackwell Publishing Ltd Oxford, UK},
	title = {Rest-frame ultraviolet-to-optical spectral characteristics of extremely metal-poor and metal-free galaxies},
	volume = {415},
	year = {2011}}

@article{fritz2006revisiting,
	author = {Fritz, Jacopo and Franceschini, Alberto and Hatziminaoglou, E},
	journal = {Monthly Notices of the Royal Astronomical Society},
	number = {3},
	pages = {767--786},
	publisher = {The Royal Astronomical Society},
	title = {Revisiting the infrared spectra of active galactic nuclei with a new torus emission model},
	volume = {366},
	year = {2006}}

@article{stalevski2012skirtor,
	author = {Stalevski, Marko},
	journal = {Bulgarian Astronomical Journal},
	number = {3},
	pages = {3},
	title = {SKIRTOR-database of modelled AGN dusty torus SEDs},
	volume = {18},
	year = {2012}}

@article{stalevski2016dust,
	author = {Stalevski, Marko and Ricci, Claudio and Ueda, Yoshihiro and Lira, Paulina and Fritz, Jacopo and Baes, Maarten},
	journal = {Monthly Notices of the Royal Astronomical Society},
	number = {3},
	pages = {2288--2302},
	publisher = {Oxford University Press},
	title = {The dust covering factor in active galactic nuclei},
	volume = {458},
	year = {2016}}

@article{yang2022fitting,
	author = {Yang, Guang and Boquien, M{\'e}d{\'e}ric and Brandt, William N and Buat, V{\'e}ronique and Burgarella, Denis and Ciesla, Laure and Lehmer, Bret D and Ma{\l}ek, Katarzyna and Mountrichas, George and Papovich, Casey and others},
	journal = {The Astrophysical Journal},
	number = {2},
	pages = {192},
	publisher = {The American Astronomical Society},
	title = {Fitting AGN/Galaxy X-Ray-to-radio SEDs with CIGALE and Improvement of the Code},
	volume = {927},
	year = {2022}}

@article{helou1985thermal,
	author = {Helou, George and Soifer, BT and Rowan-Robinson, Mo},
	journal = {Astrophysical Journal, Part 2-Letters to the Editor (ISSN 0004-637X), vol. 298, Nov. 1, 1985, p. L7-L11. NASA-supported research.},
	pages = {L7--L11},
	title = {Thermal infrared and nonthermal radio-Remarkable correlation in disks of galaxies},
	volume = {298},
	year = {1985}}

@article{draine2014infrared,
	author = {Draine, B. T. and Aniano, G. and Krause, O. and Groves, B. and Sandstrom, K. and Braun, R. and Leroy, A. and Klaas, U. and Linz, H. and Rix, H.-W. and Schinnerer, E. and Schmiedeke, A. and Walter, F.},
	journal = {ApJ},
	number = {2},
	pages = {172},
	title = {Andromeda's Dust},
	volume = {780},
	year = {2014}}

@article{evans_etal2026,
	author = {Evans, {\relax Rh}. and Gaolape, B. and Baes, Maarten and Nersesian, Angelos},
	journal = {Ap\&SS, submitted},
	title = {Choosing {CIGALE} modules and parameter grids under a memory budget: tests on {H-ATLAS} galaxies},
	year = {2026}}

@article{charlot2000simple,
	author = {Charlot, St{\'e}phane and Fall, S Michael},
	journal = {The Astrophysical Journal},
	number = {2},
	pages = {718},
	publisher = {IOP Publishing},
	title = {A Simple Model for the Absorption of Starlight by Dust inGalaxies},
	volume = {539},
	year = {2000}}

@article{chabrier2003galactic,
	author = {Chabrier, Gilles},
	journal = {Publications of the Astronomical Society of the Pacific},
	number = {809},
	pages = {763--795},
	publisher = {The University of Chicago Press},
	title = {Galactic stellar and substellar initial mass function},
	volume = {115},
	year = {2003}}

@article{salpeter1955luminosity,
	author = {Salpeter, Edwin E},
	journal = {Astrophysical Journal, vol. 121, p. 161},
	pages = {161},
	title = {The luminosity function and stellar evolution.},
	volume = {121},
	year = {1955}}

@article{bourne2016herschel,
	author = {Bourne, N and Dunne, L and Maddox, SJ and Dye, S and Furlanetto, C and Hoyos, Carlos and Smith, DJB and Eales, S and Smith, MWL and Valiante, E and others},
	journal = {Monthly Notices of the Royal Astronomical Society},
	number = {2},
	pages = {1714--1734},
	publisher = {Oxford University Press},
	title = {The Herschel-ATLAS Data Release 1--II. Multi-wavelength counterparts to submillimetre sources},
	volume = {462},
	year = {2016}}

@article{10.1093/mnras/staa3893,
	author = {Vel{\'a}zquez, Jos{\'e} A and Gurvich, Alexander B and Faucher-Gigu{\`e}re, Claude-Andr{\'e} and Bullock, James S and Starkenburg, Tjitske K and Moreno, Jorge and Lazar},
	journal = {Monthly Notices of the Royal Astronomical Society},
	month = {12},
	number = {4},
	pages = {4812-4824},
	title = {The time-scales probed by star formation rate indicators for realistic, bursty star formation histories from the FIRE simulations},
	volume = {501},
	year = {2020}}

@article{werner2004spitzer,
	author = {Werner, Michael W and Roellig, Thomas L and Low, FJ and Rieke, George H and Rieke, M and Hoffmann, WF and Young, E and Houck, JR and Brandl, B and Fazio, GG and others},
	journal = {The Astrophysical Journal Supplement Series},
	number = {1},
	pages = {1--9},
	title = {The Spitzer space telescope mission},
	volume = {154},
	year = {2004}}

@article{leitherer1999starburst99,
	author = {Leitherer, Claus and Schaerer, Daniel and Goldader, Jeffrey D and Delgado, Rosa M Gonz{\'a}lez and Robert, Carmelle and Kune, Denis Foo and De Mello, Du{\'\i}lia F and Devost, Daniel and Heckman, Timothy M},
	journal = {The Astrophysical Journal Supplement Series},
	number = {1},
	pages = {3},
	publisher = {IOP Publishing},
	title = {Starburst99: synthesis models for galaxies with active star formation},
	volume = {123},
	year = {1999}}

@article{Boquien2014,
	author = {Boquien, M. and Buat, V. and Perret, V.},
	doi = {10.1051/0004-6361/201423945},
	journal = {Astronomy \& Astrophysics},
	pages = {A72},
	title = {Impact of star formation history on the measurement of star formation rates},
	volume = {571},
	year = {2014}}

@article{draine2007infrared,
	author = {Draine, Bruce T and Li, Aigen},
	journal = {The Astrophysical Journal},
	number = {2},
	pages = {810},
	publisher = {IOP Publishing},
	title = {Infrared emission from interstellar dust. IV. The silicate-graphite-PAH model in the post-Spitzer era},
	volume = {657},
	year = {2007}}

@article{dale2014two,
	author = {Dale, Daniel A and Helou, George and Magdis, Georgios E and Armus, Lee and D{\'\i}az-Santos, Tanio and Shi, Yong},
	journal = {The Astrophysical Journal},
	number = {1},
	pages = {83},
	publisher = {IOP Publishing},
	title = {A two-parameter model for the infrared/submillimeter/radio spectral energy distributions of galaxies and active galactic nuclei},
	volume = {784},
	year = {2014}}

@article{roser2019pah,
	author = {Roser, Joseph E and Ricca, Alessandra},
	journal = {Proceedings of the International Astronomical Union},
	number = {S350},
	pages = {415--416},
	publisher = {Cambridge University Press},
	title = {PAH clusters as interstellar very small grains},
	volume = {15},
	year = {2019}}

@article{zhao2024dust,
	author = {Zhao, Roy J and Furlanetto, Steven R},
	journal = {Journal of Cosmology and Astroparticle Physics},
	number = {09},
	pages = {018},
	publisher = {IOP Publishing},
	title = {Dust in high-redshift galaxies: reconciling UV attenuation and IR emission},
	volume = {2024},
	year = {2024}}

@article{salim2016galex,
	author = {Salim, Samir and Lee, Janice C and Janowiecki, Steven and Da Cunha, Elisabete and Dickinson, Mark and Boquien, M{\'e}d{\'e}ric and Burgarella, Denis and Salzer, John J and Charlot, St{\'e}phane},
	journal = {The Astrophysical Journal Supplement Series},
	number = {1},
	pages = {2},
	publisher = {IOP Publishing},
	title = {GALEX--SDSS--WISE Legacy Catalog (GSWLC): Star formation rates, stellar masses, and dust attenuations of 700,000 low-redshift galaxies},
	volume = {227},
	year = {2016}}

@article{Conroy_2013,
	author = {Conroy, Charlie},
	doi = {10.1146/annurev-astro-082812-141017},
	issn = {1545-4282},
	journal = {Annual Review of Astronomy and Astrophysics},
	month = {aug},
	number = {1},
	pages = {393--455},
	publisher = {Annual Reviews},
	title = {Modeling the Panchromatic Spectral Energy Distributions of Galaxies},
	url = {http://dx.doi.org/10.1146/annurev-astro-082812-141017},
	volume = {51},
	year = {2013}}

@article{ferland20132013,
	author = {Ferland, Gary J and Porter, RL and Van Hoof, PAM and Williams, RJR and Abel, NP and Lykins, ML and Shaw, Gargi and Henney, William J and Stancil, PC},
	journal = {Revista mexicana de astronom{\'\i}a y astrof{\'\i}sica},
	number = {1},
	pages = {137--163},
	publisher = {Instituto de Astronom{\'\i}a},
	title = {The 2013 release of cloudy},
	volume = {49},
	year = {2013}}

@article{ferland1998cloudy,
	author = {Ferland, GJ and Korista, KT and Verner, DA and Ferguson, JW and Kingdon, JB and Verner, EM},
	journal = {Publications of the Astronomical Society of the Pacific},
	number = {749},
	pages = {761},
	publisher = {IOP Publishing},
	title = {CLOUDY 90: numerical simulation of plasmas and their spectra},
	volume = {110},
	year = {1998}}

@article{lintott2011galaxy,
	author = {Lintott, Chris and Schawinski, Kevin and Bamford, Steven and Slosar, An{\v{z}}e and Land, Kate and Thomas, Daniel and Edmondson, Edd and Masters, Karen and Nichol, Robert C and Raddick, M Jordan and others},
	journal = {Monthly Notices of the Royal Astronomical Society},
	number = {1},
	pages = {166--178},
	publisher = {OUP},
	title = {Galaxy Zoo 1: data release of morphological classifications for nearly 900 000 galaxies},
	volume = {410},
	year = {2011}}

@ARTICLE{Meurer1999,
  author  = {Meurer, Gerhardt R. and Heckman, Timothy M. and Calzetti, Daniela},
  title   = {Dust Absorption and the Ultraviolet Luminosity Density at z $\approx$ 3 as Calibrated by Local Starburst Galaxies},
  journal = {The Astrophysical Journal},
  year    = {1999},
  volume  = {521},
  pages   = {64}
}

@ARTICLE{Buat2005,
  author  = {Buat, V. and others},
  title   = {Dust Attenuation in the Nearby Universe: A Comparison between Galaxies Selected in the Ultraviolet and in the Far-Infrared},
  journal = {The Astrophysical Journal},
  year    = {2005},
  volume  = {619},
  pages   = {L51}
}

@ARTICLE{Noll2009,
  author  = {Noll, S. and Burgarella, D. and Giovannoli, E. and Buat, V. and Marcillac, D. and Mu{\~n}oz-Mateos, J. C.},
  title   = {Analysis of galaxy spectral energy distributions from far-UV to far-IR with CIGALE: studying a SINGS test sample},
  journal = {Astronomy \& Astrophysics},
  year    = {2009},
  volume  = {507},
  pages   = {1793}
}

@ARTICLE{SalimNarayanan2020,
  author  = {Salim, Samir and Narayanan, Desika},
  title   = {The Dust Attenuation Law in Galaxies},
  journal = {Annual Review of Astronomy and Astrophysics},
  year    = {2020},
  volume  = {58},
  pages   = {529}
}

@ARTICLE{Casey2012,
  author  = {Casey, Caitlin M.},
  title   = {Far-infrared spectral energy distribution fitting for galaxies near and far},
  journal = {Monthly Notices of the Royal Astronomical Society},
  year    = {2012},
  volume  = {425},
  pages   = {3094}
}

@article{KennicuttEvans2012,
  author  = {Kennicutt, Jr., Robert C. and Evans, II, Neal J.},
  title   = {Star Formation in the Milky Way and Nearby Galaxies},
  journal = {Annual Review of Astronomy and Astrophysics},
  year    = {2012},
  volume  = {50},
  pages   = {531--608},
  doi     = {10.1146/annurev-astro-081811-125610}
}

@ARTICLE{Bennett2014,
  author  = {Bennett, C. L. and Larson, D. and Weiland, J. L. and Hinshaw, G.},
  title   = {The 1\% Concordance Hubble Constant},
  journal = {The Astrophysical Journal},
  year    = {2014},
  volume  = {794},
  pages   = {135}
}

@ARTICLE{Martin2005, author={{Martin}, D. Christopher and others}, title={The Galaxy Evolution Explorer: A Space Ultraviolet Survey Mission}, journal={ApJ}, year=2005, volume=619, pages={L1}}

@ARTICLE{York2000, author={{York}, Donald G. and others}, title={The Sloan Digital Sky Survey: Technical Summary}, journal={AJ}, year=2000, volume=120, pages={1579}}

@ARTICLE{Edge2013, author={{Edge}, A. and {Sutherland}, W. and {Kuijken}, K. and others}, title={The VISTA Kilo-degree Infrared Galaxy (VIKING) Survey: Bridging the Gap between Low and High Redshift}, journal={The Messenger}, year=2013, volume=154, pages={32}}

@ARTICLE{Wright2010, author={{Wright}, Edward L. and others}, title={The Wide-field Infrared Survey Explorer (WISE): Mission Description and Initial On-orbit Performance}, journal={AJ}, year=2010, volume=140, pages={1868}}

@ARTICLE{Driver2016, author={{Driver}, S. P. and others}, title={GAMA Panchromatic Data Release (far-UV-far-IR) and the low-z energy budget}, journal={MNRAS}, year=2016, volume=455, pages={3911}}

@ARTICLE{stern2012mid,
  author  = {{Stern}, D. and {Assef}, R.~J. and {Benford}, D.~J. and {Blain}, A. and {Cutri}, R. and {Dey}, A. and {Eisenhardt}, P. and {Griffith}, R.~L. and {Jarrett}, T.~H. and {Lake}, S. and others},
  title   = "{Mid-infrared Selection of Active Galactic Nuclei with the Wide-Field Infrared Survey Explorer. I. Characterizing WISE-selected Active Galactic Nuclei in COSMOS}",
  journal = {\apj},
  year    = 2012,
  volume  = 753,
  number  = 1,
  pages   = {30},
  doi     = {10.1088/0004-637X/753/1/30}
}

@ARTICLE{schreiber2018egg,
  author  = {{Schreiber}, C. and {Elbaz}, D. and {Pannella}, M. and {Ciesla}, L. and {Wang}, T. and {Franco}, M.},
  title   = "{Dust temperature and mid-to-total infrared color distributions for star-forming galaxies at 0 < z < 4}",
  journal = {\aap},
  year    = 2018,
  volume  = 609,
  pages   = {A30},
  doi     = {10.1051/0004-6361/201731506}
}

@article{evans_memory2026,
  author  = {Evans, {\relax Rh}. and Gaolape, B.},
  title   = {Running {CIGALE} on large model grids with limited memory: the roles of cores and blocks},
  journal = {Ap\&SS, submitted},
  year    = 2026
}

@MISC{cutri2013allwise,
  author       = {{Cutri}, R.~M. and others},
  title        = "{Explanatory Supplement to the AllWISE Data Release Products}",
  howpublished = {IPAC/Caltech},
  year         = 2013
}

@ARTICLE{jarrett2011spitzer,
  author       = {{Jarrett}, T.~H. and others},
  title        = "{The Spitzer-WISE Survey of the Ecliptic Poles}",
  journal      = {ApJ},
  year         = 2011,
  volume       = 735,
  pages        = {112}
}

@ARTICLE{smith2013isothermal,
  author  = {{Smith}, D.~J.~B. and others},
  title   = "{Isothermal dust models of Herschel-ATLAS galaxies}",
  journal = {MNRAS},
  year    = 2013,
  volume  = 436,
  pages   = {2435},
  doi     = {10.1093/mnras/stt1737}
}

@ARTICLE{amblard2010herschel,
  author  = {{Amblard}, A. and others},
  title   = "{Herschel-ATLAS: Dust temperature and redshift distribution of SPIRE and PACS detected sources using submillimetre colours}",
  journal = {A\&A},
  year    = 2010,
  volume  = 518,
  pages   = {L9},
  doi     = {10.1051/0004-6361/201014586}
}

@ARTICLE{symeonidis2013herschel,
  author  = {{Symeonidis}, M. and others},
  title   = "{The Herschel census of infrared SEDs through cosmic time}",
  journal = {MNRAS},
  year    = 2013,
  volume  = 431,
  pages   = {2317},
  doi     = {10.1093/mnras/stt330}
}

\appendix

\section{\textsc{cigale} configuration}
\label{app:cigale}

\begin{figure*}
	\centering
	\includegraphics[width=\textwidth]{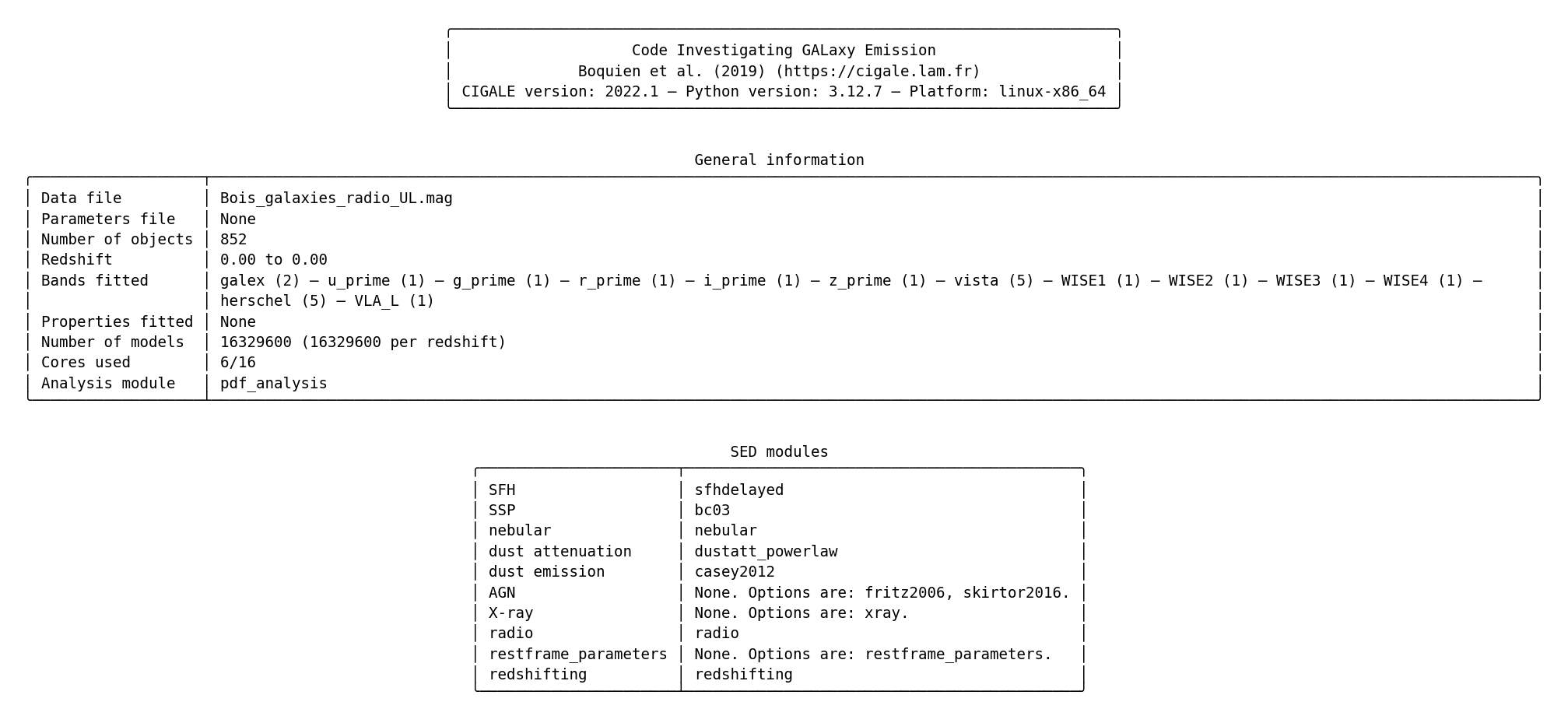}
	\caption{General information from the \textsc{cigale} run of the final fits (\texttt{pcigale check}; v2022.1 on the Linux mini PC, 16\,329\,600 models, six cores of the 16 threads). The fit was made for the 852 galaxies of the Thesis sample, and the \Nsamp\ galaxies of this paper are taken from it. All galaxies are fitted at $z=0$ (Section~\ref{sec:redshift}), and 714 of the 1.4\,GHz flux densities are upper limits.}
	\label{fig:general info on models}
	\par\smallskip{\small Alt text: Screenshot of the text summary printed by \textsc{cigale}: the data file, 852 objects, the redshift fixed at 0, the 22 bands fitted, 16\,329\,600 models, 6 of 16 cores used, and the modules \texttt{sfhdelayed}, \texttt{bc03}, \texttt{nebular}, \texttt{dustatt\_powerlaw}, \texttt{casey2012}, \texttt{radio} and \texttt{redshifting}.}
\end{figure*}

Fig.~\ref{fig:general info on models} shows the general information from the \textsc{cigale} run, and Table~\ref{tab:cigale_parameters} gives the parameters and values specified for each module.

\begin{table*}
    \centering
    \caption{Parameters and values specified for the \textsc{cigale} modules used (listed in Table~\ref{tab:modules}), for the final fits ($16\,329\,600$ models per galaxy). Only active configuration entries are included. The 1.4\,GHz flux densities of 714 of the \Nsamp\ galaxies are entered as upper limits (Section~\ref{sec:data}).}
    \label{tab:cigale_parameters}
    \small
    \renewcommand{\arraystretch}{1.1}
    \begin{tabularx}{\textwidth}{@{}lX@{}}
        \toprule
        Parameter & Value(s) \\
        \midrule
        \multicolumn{2}{@{}l}{\textbf{Star formation history:} \texttt{sfhdelayed}} \\
        \texttt{tau\_main} & 100, 500, 750, 1000, 2000, 4000, 8000, 15000 \\
        \texttt{age\_main} & 1000, 1500, 2000, 3000, 4000, 6000, 8000, 10000, 12000 \\
        \texttt{tau\_burst} & 50.0 \\
        \texttt{age\_burst} & 20 \\
        \texttt{f\_burst} & 0.0 \\
        \texttt{sfr\_A} & 1.0 \\
        \texttt{normalise} & \texttt{True} \\
        \midrule
        \multicolumn{2}{@{}l}{\textbf{Single stellar population:} \texttt{bc03}} \\
        \texttt{imf} & 0 (Salpeter) \\
        \texttt{metallicity} & 0.008, 0.02, 0.05 \\
        \texttt{separation\_age} & 10 \\
        \midrule
        \multicolumn{2}{@{}l}{\textbf{Nebular:} \texttt{nebular}} \\
        \texttt{logU} & $-2.0$ \\
        \texttt{zgas} & 0.02 \\
        \texttt{ne} & 100 \\
        \texttt{f\_esc} & 0.0 \\
        \texttt{f\_dust} & 0.0 \\
        \texttt{lines\_width} & 300.0 \\
        \texttt{emission} & \texttt{True} \\
        \midrule
        \multicolumn{2}{@{}l}{\textbf{Dust attenuation:} \texttt{dustatt\_powerlaw}} \\
        \texttt{Av\_young} & 0.25, 0.5, 1.0, 1.5, 2.0, 3.0, 4.0, 5.0, 6.0, 8.0 \\
        \texttt{Av\_old\_factor} & 0.1, 0.2, 0.3, 0.45, 0.6, 0.8, 1.0 \\
        \texttt{uv\_bump\_wavelength} & 217.5 \\
        \texttt{uv\_bump\_width} & 35.0 \\
        \texttt{uv\_bump\_amplitude} & 0.0, 1.5, 3.0 \\
        \texttt{powerlaw\_slope} & $-0.7$ \\
        \texttt{filters} & \texttt{NUV \& FUV \& u\_prime \& g\_prime \& r\_prime} \\
        \midrule
        \multicolumn{2}{@{}l}{\textbf{Dust emission:} \texttt{casey2012}} \\
        \texttt{temperature} & 17.5, 20.0, 22.5, 25.0, 27.5, 30.0, 32.5, 35.0, 37.5, 40.0, 45.0, 50.0 \\
        \texttt{beta} & 1.0, 1.5, 1.8, 2.1, 2.5 \\
        \texttt{alpha} & 1.5, 1.75, 2.0, 2.25, 2.5, 3.0 \\
        \midrule
        \multicolumn{2}{@{}l}{\textbf{Radio:} \texttt{radio}} \\
        \texttt{qir\_sf} & 2.58 \\
        \texttt{alpha\_sf} & 0.8 \\
        \texttt{R\_agn} & 10 \\
        \texttt{alpha\_agn} & 0.7 \\
        \bottomrule
    \end{tabularx}
\end{table*}

\begin{figure*}
	\includegraphics[width=0.85\textwidth]{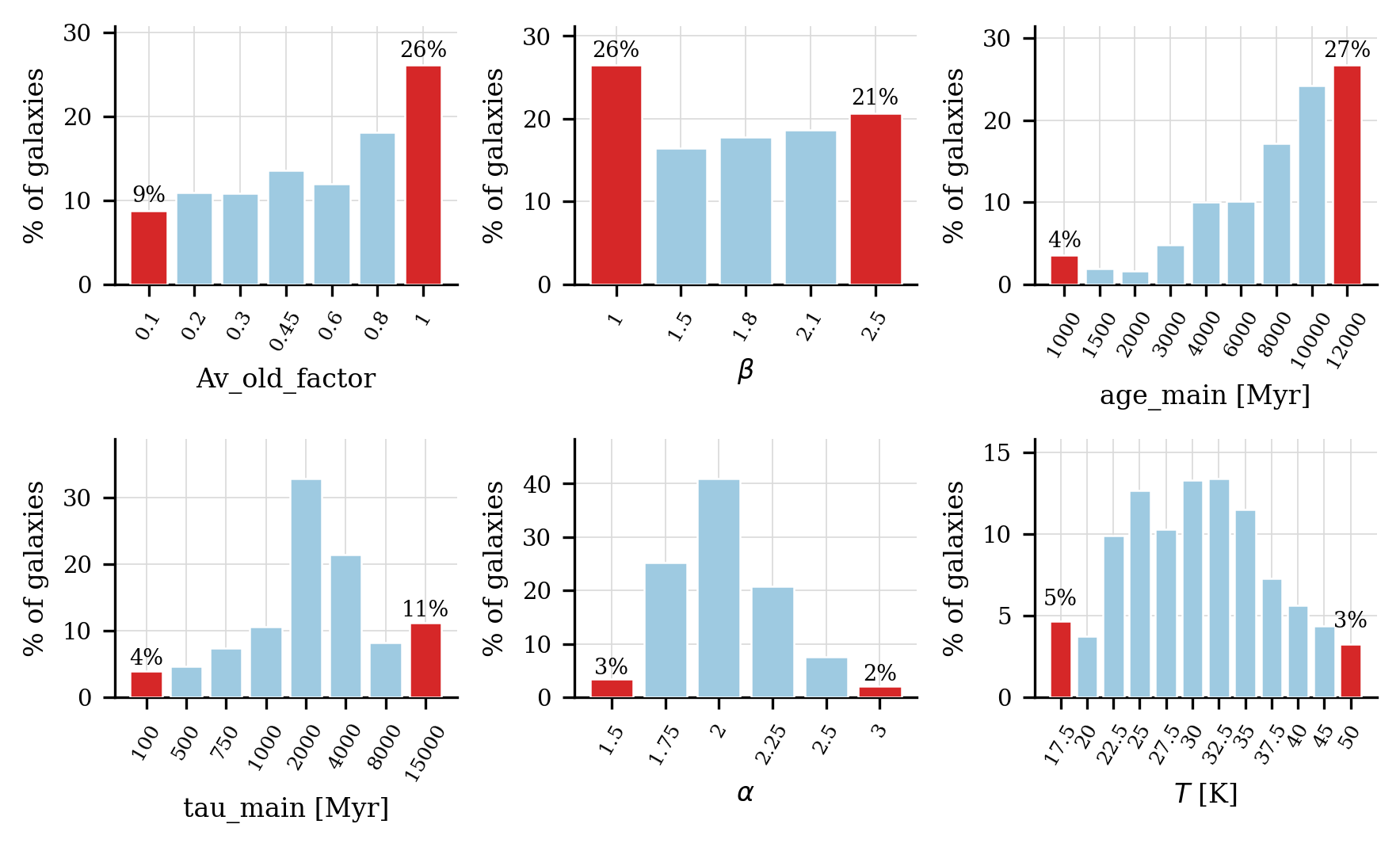}
	\caption{Distribution of the best-fitting values of six model parameters among the \Nsamp\ galaxies, as the percentage of the sample at each value of the grid (Table~\ref{tab:cigale_parameters}). The first and last values of each grid are highlighted, with their percentages.}
	\label{fig:grid_edges}
	\par\smallskip{\small Alt text: Six bar charts of the percentage of galaxies at each grid value of \texttt{Av\_old\_factor}, $\beta$, \texttt{age\_main}, \texttt{tau\_main}, $\alpha$ and the dust temperature. The first and last bars of each chart are red and labelled: 9 and 26 per cent for \texttt{Av\_old\_factor}, 26 and 21 per cent for $\beta$, 4 and 27 per cent for the age, 4 and 11 per cent for $\tau$, 3 and 2 per cent for $\alpha$, and 5 and 3 per cent for the temperature.}
\end{figure*}

\bsp
\label{lastpage}
\end{document}